\documentclass[pdflatex,sn-mathphys-num]{sn-jnl_mod}

\usepackage{amsmath,amssymb,amsfonts}
\usepackage{amsthm}
\usepackage{orcidlink}
\usepackage{graphicx}
\usepackage{booktabs}
\usepackage[parfill]{parskip}
\usepackage{esint}
\usepackage{multirow}
\usepackage{changes}

\newcommand{\bra}[1]{{\left\langle #1\right|}}
\newcommand{\ket}[1]{{\left| #1 \right\rangle}}

\begin{document}

\title{Extraction of DVCS amplitudes off the nucleon}

\author[1,2]{C.~Mezrag\,\orcidlink{0000-0001-8678-4085}}
\author[3]{P.~Sznajder\,\orcidlink{0000-0002-2684-803X}}
\author[3]{J.~Wagner\,\orcidlink{0000-0001-8335-7096}}

\affil[1]{Universit\'e Paris-Saclay - CEA - IRFU, F-91191 Gif-sur-Yvette, France}
\affil[2]{AIDAS, CEA, F-91191, Gif-sur-Yvette, France}

\affil[3]{National Centre for Nuclear Research (NCBJ), Pasteura 7, 02-093 Warsaw, Poland}

\abstract{
This work presents a modern extraction of deeply virtual Compton scattering (DVCS) amplitudes off the proton and off the neutron through a global analysis of experimental data. These amplitudes serve as a crucial intermediate quantity linking generalized parton distributions (GPDs) to experimental observables. The analysis relies on a novel extraction framework recently integrated into the PARTONS software ecosystem. We employ two distinct modelling approaches for the DVCS amplitudes, both utilizing machine learning techniques: one model-agnostic and the other theory-augmented. For the first time, we incorporate the extraction of helicity-flip amplitudes, which are particularly sensitive to higher-twist effects. Furthermore, we compare our extracted amplitudes to timelike Compton scattering (TCS) data using the established relations between DVCS and TCS, serving as an important test of universality of GPDs. We also extract the DVCS subtraction constant, for which \textit{ab initio} predictions from lattice QCD have been obtained. This allows us to confront these predictions with DVCS data for the first time, initiating a new virtuous cycle between first-principles theory and phenomenology.
}


\maketitle

\section{Introduction}
\label{sec:intro}

Generalised parton distributions (GPDs) were introduced in the 1990s~\cite{Mueller:1998fv,Ji:1996ek,Radyushkin:1997ki} and have been extensively studied since then for three main reasons. 
Firstly, they allow one to describe deep exclusive processes such as deeply virtual Compton scattering (DVCS)~\cite{Ji:1996nm,Ji:1998xh,Collins:1998be} and deeply virtual meson production (DVMP)~\cite{Collins:1996fb}. 
Secondly, they offer the ability to draw a 3D picture of the number density of quarks and gluons within hadrons~\cite{Burkardt:2000za,Diehl:2002he}. 
This feature is a highly desirable generalisation of projections that were previously achievable from parton distribution functions (PDFs) and electromagnetic form factors (EFFs). 
Thirdly, GPDs provide the sole experimental access known today to the energy-momentum tensor (EMT) of hadrons~\cite{Ji:1996ek,Polyakov:2002yz}, allowing one to extract the associated form factors.

Of all processes sensitive to GPDs, DVCS has certainly been the most studied one experimentally, as data are available from H1~\cite{H1:2001nez,Aktas:2005ty,H1:2007vrx,H1:2009wnw}, ZEUS~\cite{ZEUS:2003pwh,ZEUS:2008hcd}, CLAS~\cite{CLAS:2001wjj,CLAS:2006krx,Girod:2007aa,CLAS:2008ahu,CLAS:2015bqi,Jo:2015ema,CLAS:2018bgk,CLAS:2022syx,CLAS:2024qhy}, HERMES~\cite{HERMES:2001bob,HERMES:2006pre,HERMES:2008abz,HERMES:2009cqe,HERMES:2010dsx,HERMES:2011bou,HERMES:2012gbh}, Hall A~\cite{MunozCamacho:2006hx,Defurne:2015kxq,Defurne:2017paw,JeffersonLabHallA:2022pnx,Benali:2020vma}, and COMPASS~\cite{COMPASS:2018pup} collaborations. 
It has also been extensively studied theoretically. 
Indeed, the coefficient function is now known at next-to-next-to-leading order (NNLO)~\cite{Braun:2022bpn} in the strong running coupling, $\alpha_S$, and at next-to-next-to-leading power (NNLP)~\cite{Braun:2012bg,Braun:2012hq,Braun:2014sta,Martinez-Fernandez:2025gub,Braun:2025xlp}. 
In parallel, the GPD evolution equations have been computed at three loops in the non-singlet case~\cite{Braun:2017cih} and at two loops in the singlet case~\cite{Belitsky:1999fu,Braun:2019qtp}. 
Modern numerical implementations have also been developed~\cite{Bertone:2017gds,Bertone:2022frx,Freese:2024ypk}. 
The dispersive formalism, connecting the real and imaginary parts of the Compton form factors, has been extended to NLO and NLP~\cite{Diehl:2007jb,Dutrieux:2024bgc,Martinez-Fernandez:2025rcg,Martinez-Fernandez:2025jvk}.

Despite these efforts to achieve a level of precision in DVCS comparable to that of the PDF and TMD communities, phenomenological extractions face a major obstacle known as shadow GPDs~\cite{Bertone:2021yyz,Moffat:2023svr}. 
These shadow GPDs are eigenfunctions associated with the vanishing eigenvalues of the coefficient function. 
Mathematically, they form the kernel of the coefficient function acting as an operator, thereby spanning an \textit{a priori} infinite-dimensional functional subspace. 
As the precision of both experimental data and perturbative QCD computations improves, this functional ambiguity emerges as the dominant source of theoretical uncertainty. 
Fortunately, the enforcement of rigorous theoretical constraints, such as polynomiality and positivity, combined with multi-channel analyses incorporating processes like double DVCS~\cite{Belitsky:2002tf,Guidal:2002kt,Deja:2023ahc,Alvarado:2026ggy}, is expected to significantly mitigate this uncertainty in future extractions~\cite{Dutrieux:2021wll}.

Lattice QCD has also been advocated as a potential way to bypass the deconvolution problem of DVCS by exploiting simulated data together with experimental data~\cite{Riberdy:2023awf,Cichy:2024afd,Guo:2025muf}. 
This has been made possible thanks to the formalisms of large momentum effective theory (LaMET, or quasi-distributions)~\cite{Ji:2013dva} and short-distance factorisation (SDF, or pseudo-distributions)~\cite{Radyushkin:2017cyf} (see also other approaches such as~\cite{Braun:2007wv,Ma:2014jla}). 
Without entering into details, lattice data are connected to GPDs through two main ingredients: a perturbative coefficient function, called the matching kernel, and a Fourier transform, allowing one to go from the position space where lattice data are produced to the momentum space where GPDs are defined. 
These matching kernels differ from those describing exclusive processes, thereby offering complementary insights into hadron structure.
However, extracting GPDs from the lattice faces its own challenges, including a distinct deconvolution problem currently under debate~\cite{Dutrieux:2025jed,Chen:2025cxr,Dutrieux:2025axb,Xiong:2025obq,Chu:2025jsi}. 
Nevertheless, attempts to extract GPDs from the lattice have been performed using both formalisms, alongside other approaches~\cite{Hannaford-Gunn:2024aix,Ding:2024saz,Chu:2025kew,Dutrieux:2026grg}.

In spite of the aforementioned difficulties, several extractions have been performed in the past, employing different strategies. 
Notable examples include parametric fits such as the Goloskokov-Kroll (GK)~\cite{Goloskokov:2007nt}, Vanderhaeghen-Guichon-Guidal (VGG)~\cite{Vanderhaeghen:1999xj}, and Liuti-Hernandez-Goldstein~\cite{Goldstein:2010gu} models in momentum space, efforts regarding the parametrisation of conformal moments~\cite{Kumericki:2009uq,Kumericki:2013br,Kumericki:2015lhb}, and non-parametric attempts~\cite{Kumericki:2011rz}. 
Beyond these pioneering attempts, additional developments can be found in the literature~\cite{Cuic:2020iwt,Dupre:2016mai, Hashamipour:2021kes,Guo:2023ahv,Guo:2025muf,Xu:2026lko}, including a major milestone achieved in Ref.~\cite{Cuic:2023mki}, where both DVCS and DVMP are described at NLO precision using the same model in conformal space, confirming the universal nature of GPDs.
Previously, using the PARTONS framework~\cite{Berthou:2015oaw}, we completed both a parametric~\cite{Moutarde:2018kwr} and a non-parametric~\cite{Moutarde:2019tqa} description of DVCS observables. 

We believe it is timely to revisit the analyses performed in 2018 for the following reasons:
\begin{itemize}
    \item new JLab 12 GeV data have become available, significantly extending the kinematic coverage.
    \item our understanding of the underlying physics has advanced, particularly regarding the deconvolution problem.
    \item in the coming years, lattice QCD computations may reach precision levels competitive with phenomenological analyses.
\end{itemize}
Consequently, the GPD field has reached a level of maturity where sources of uncertainty must be rigorously controlled, evaluated, and propagated.

To achieve this objective, the present work introduces a new extraction of DVCS amplitudes from available proton and neutron data, utilizing a recently developed framework within the PARTONS software ecosystem. We employ two distinct modelling strategies for the constrained quantities. 
The first approach relies solely on artificial neural networks (ANNs) and can be considered ``unbiased''\footnote{Naturally, no extraction is truly unbiased, as the network architecture dictates how the solution space is explored.} in the sense that no explicit theoretical constraints are imposed during the fit. 
In contrast, the second approach incorporates physics-driven insights regarding the DVCS amplitudes, notably enforcing the dispersion relations that connect their real and imaginary parts. This study is complemented by a comparison of the obtained results against available TCS data. This serves as an important universality test and, though not directly, demonstrates the statistical impact of the TCS measurements. Furthermore, we compare our extracted subtraction constant with selected predictions from lattice QCD.

The extraction of CFFs, the most fundamental quantities carrying information about GPDs that can be unambiguously determined from data, should be viewed as a crucial intermediate step toward the full determination of GPDs. Although solving the deconvolution problem to obtain a meaningful GPD extraction from these amplitudes is left for future work, extracting CFFs first allows the preparatory work for determining GPDs to be performed without rerunning computationally expensive fits to the experimental data. Moreover, the extracted CFFs can be regarded as a refined distillation of the experimental data collected over the last two decades, providing an ideal tool for impact studies, as demonstrated in Refs.~\cite{Grocholski:2019pqj,Anderle:2021wcy,Dutrieux:2021ehx}. Finally, the subtraction constant appearing in the dispersion relations offers direct access to elements of the hadronic energy-momentum tensor form factors.

The structure of this article is as follows: in Sect.~\ref{sec:theory}, we present the theoretical foundations of this work, focusing on the elements of the GPD formalism that are crucial for understanding this study. In Sect.~\ref{sec:extraction}, we introduce a novel extraction framework, utilized here for the first time, which will serve as a reference for future works. In Sect.~\ref{sec:ansatz}, we present two Ansätze and detail the methodology, including our approach to addressing data sparsity in certain kinematic domains, the regularization methods employed, the procedures for fixing hyperparameters and minimization, and the propagation of uncertainties. Section~\ref{sec:result} begins with a presentation of the dataset used and an evaluation of the fit performance. Subsequently, the extracted quantities are discussed, including an analysis of the subtraction constant and a comparison against TCS data. The summary is provided in Sect.~\ref{sec:conclusions}.


\section{Theory framework}
\label{sec:theory}

We consider the exclusive electroproduction of a real photon,
\begin{equation} 
    e(k) + p(p) \to e(k') + p(p') + \gamma(q')\,, 
    \label{eq:theory:DVCS}
\end{equation} 
where $k$ ($k'$) and $p$ ($p'$) are the four-momenta of the incoming (outgoing) lepton and nucleon, respectively, and $q'$ is the four-momentum of the produced real photon. We define the average momentum $P=(p+p')/2$ and the momentum transfers as 
\begin{equation} 
    q = k-k', \qquad \Delta = p'-p = q-q'\,. 
\end{equation} 

The kinematics of the process in Eq.~\eqref{eq:theory:DVCS} are characterized by the following Lorentz-invariant variables:
\begin{equation} 
    Q^2=-q^2, \qquad x_B=\frac{Q^2}{2p\cdot q}, \qquad t=\Delta^2\,, 
\end{equation} 
along with the azimuthal angle $\phi$ between the leptonic and hadronic scattering planes. For the latter, we adopt the standard Trento convention~\cite{Bacchetta:2004jz}. In terms of these variables, the four-fold differential cross section is expressed as
\begin{equation}
    \frac{d^4\sigma^{b,c}_{t}}{dx_B\,dQ^2\,dt\,d\phi}\,,
    \label{eq:electroproductionCrossSection}
\end{equation}
where $b$, $c$, and $t$ denote the beam helicity, beam charge, and target polarization state, respectively. The observables used in this analysis, such as measured cross sections, cross-section differences, and asymmetries, are constructed from combinations of these polarization- and charge-dependent differential cross sections (see, e.g., Ref.~\cite{Moutarde:2018kwr}).

The scattering amplitude for the exclusive electroproduction of a real photon receives contributions from both the DVCS and the Bethe-Heitler (BH) processes,
\begin{equation}
    \mathcal{T} = \mathcal{T}_{\mathrm{DVCS}} + \mathcal{T}_{\mathrm{BH}}\,.
\end{equation}
Consequently, the squared amplitude is given by
\begin{equation}
    |\mathcal{T}|^2 = |\mathcal{T}_{\mathrm{BH}}|^2 + |\mathcal{T}_{\mathrm{DVCS}}|^2 + \mathcal{I}\,,
\end{equation}
where
\begin{equation}
    \mathcal{I} = \mathcal{T}_{\mathrm{DVCS}} \mathcal{T}_{\mathrm{BH}}^\ast + \mathcal{T}_{\mathrm{DVCS}}^\ast \mathcal{T}_{\mathrm{BH}}
\end{equation}
is the interference term. The BH amplitude is fully computable from the elastic form factors (EFFs) of the nucleon. In this analysis we use the parametrisation of EFFs reported in Ref.~\cite{Ye:2017gyb}. The hadronic structure governing the DVCS amplitude is encoded in the Compton tensor $T^{\mu\nu}$ associated with the subprocess
\begin{equation}
    \gamma^\ast(q) + p(p) \to \gamma(q') + p(p')\,.
\end{equation}
It is convenient to decompose this tensor into photon-helicity amplitudes. Following Refs.~\cite{Belitsky:2001ns,Belitsky:2012ch}, we define
\begin{equation}
    \mathcal{T}^{\mathrm{DVCS}}_{\lambda\lambda'}
    =
    \varepsilon_\mu^\ast(q',\lambda')\,
    T^{\mu\nu}\,
    \varepsilon_\nu(q,\lambda)\,,
\end{equation}
where $\varepsilon_\nu(q,\lambda)$ and $\varepsilon_\mu^\ast(q',\lambda')$ are the polarization vectors of the incoming virtual and outgoing real photons, respectively. The indices $\lambda \in \{+,0,-\}$ and $\lambda' \in \{+,-\}$ denote the corresponding photon helicities.

Crucially to this work, for each photon-helicity transition, the corresponding amplitude, $\mathcal{T}^{\mathrm{DVCS}}_{\lambda\lambda'}$, can be parameterized in terms of a set of CFFs, 
\begin{align} 
\mathcal{F}^{\lambda\lambda'} \in \left\{ \mathcal{H}^{\lambda\lambda'}, \mathcal{E}^{\lambda\lambda'}, \widetilde{\mathcal{H}}^{\lambda\lambda'}, \widetilde{\mathcal{E}}^{\lambda\lambda'} \right\}\,. 
\end{align}
Parity symmetry relates the remaining helicity amplitudes, thus, it is sufficient to consider only the three independent sectors:
\begin{align} 
    \mathcal{F}^{++}\,, \qquad \mathcal{F}^{0+}\,,  \qquad \mathcal{F}^{+-} \,, 
\end{align} 
which correspond, respectively, to photon-helicity conservation, a longitudinal-to-transverse transition, and a transverse transition. 

It is important to emphasize that throughout this work, we adopt the definition of CFFs from the Belitsky-Müller-Ji (BMJ) formalism~\cite{Belitsky:2012ch}, as implemented in the PARTONS framework~\cite{Berthou:2015oaw}. Alternative definitions exist, with differences typically stemming from the choice of reference frame, which manifests as kinematic higher-twist corrections. The most prominent example of such an alternative approach is the Braun-Manashov-Pirnay (BMP) formalism~\cite{Braun:2014sta}. Note that the exact translations between the BMJ and BMP definitions are well established.

Within collinear factorization, the CFFs are expressed as convolutions of perturbatively calculable coefficient functions with GPDs. Although in the present analysis the deconvolution of CFFs into GPDs is not attempted, it is still instructive to see how both quantities are related:
\begin{align}
    \mathcal{F}^{\lambda\lambda'}(\xi,t,Q^2)
    =
    \sum_i
    \int_{-1}^{1} dx\,
    C_i^{\lambda\lambda'}
    \left(
        x,\xi,\frac{Q^2}{\mu^2},\alpha_s(\mu^2)
    \right)
    F_i(x,\xi,t,\mu^2)
    +
    \mathcal{O}\left(\frac{1}{Q}\right)\,.
    \label{eq:CFFconvolution}
\end{align}
Here, $i$ runs over the relevant quark and gluon channels. The terms $C_i^{\lambda\lambda'}$ and $F_i$ represent the coefficient functions and GPDs, respectively, while $\xi$ denotes the skewness variable and $\mu^2$ is the perturbative scale (typically chosen as $Q^2$ for DVCS). When extracting CFFs from experimental data, we use the following relation:
\begin{equation}
    \xi = \frac{x_B}{2-x_B} \,,
\end{equation}
which is consistent with the BMJ definition of CFFs.

The DVCS amplitudes are complex functions whose real and imaginary parts can be connected through dispersion relations that can be derived from analyticity and crossing symmetries~\cite{Teryaev:2005uj,Anikin:2007yh,Diehl:2007jb,Braun:2014sta,Dutrieux:2024bgc,Martinez-Fernandez:2025rcg}. For the CFF $\mathcal{H}^{++}$, one has
\begin{align}
    \label{eq:DVCSDispersionRelation}
    \mathcal{S}(t,Q^2) = \mathrm{Re} \mathcal{H}^{++}(\xi,t,Q^2) - \frac{2}{\pi} \fint_0^1 \frac{x\, \mathrm{Im} \mathcal{H}^{++}(x,t,Q^2)}{(\xi-x)(\xi +x)} d x \,.
\end{align}
where $\fint$ denotes the principal value integral, and $\mathcal{S}$ is the subtraction constant, which plays an important role in the determination of the energy-momentum tensor form factors~\cite{Polyakov:2002yz,Martinez-Fernandez:2025jvk}. Similar relations exist for other helicity-conserving CFFs, where the subtraction constant has exactly the same magnitude but the opposite sign for the CFF $\mathcal{E}^{++}$, and it completely vanishes for $\mathcal{\widetilde{H}}^{++}$ and $\mathcal{\widetilde{E}}^{++}$. For non-helicity-conserving CFFs, such relations have not yet been derived.

\section{Extraction framework}
\label{sec:extraction}

\begin{figure}[!ht]
    \centering
    \includegraphics[width=\textwidth]{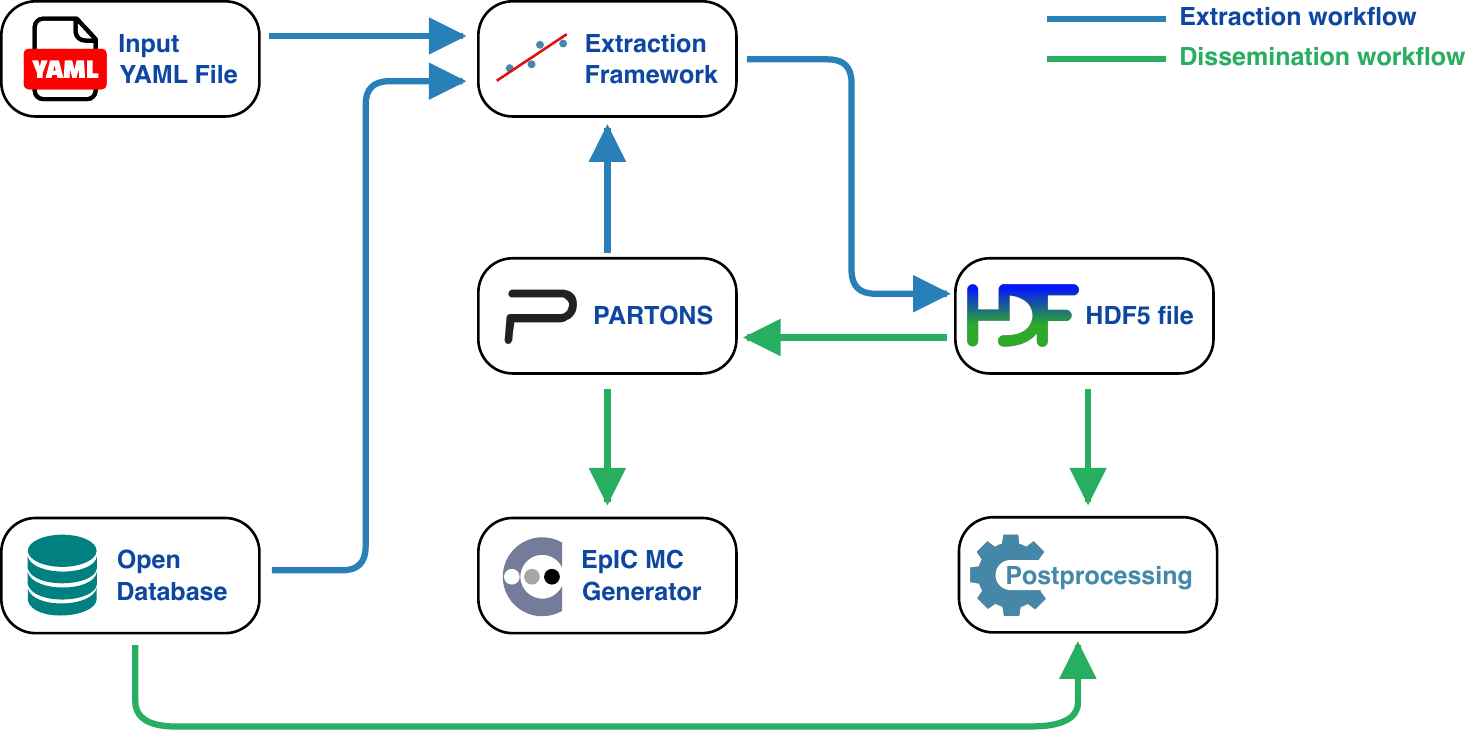}
    \caption{Schematic illustration of the extraction and dissemination workflows. Further details are provided in the text.}
    \label{fig:extraction:workflow}
\end{figure}

The general structure of the extraction framework used in this work is depicted in Fig.~\ref{fig:extraction:workflow}. The framework is 
written in C++ and is based on PARTONS~\cite{Berthou:2015oaw}, an open-source library for studying GPDs. PARTONS can be used to evaluate observables for exclusive processes from models of GPDs, as well as directly from parameterizations of amplitudes (CFFs for DVCS), the latter being crucial for this work. Furthermore, the extraction framework utilizes core PARTONS functionalities, such as the registry mechanism and logger. This allows for full integration between the two software packages, thereby inheriting key features that support, among other things, a highly modular structure. 

The framework is configured via a YAML input file. This format is a human-readable data serialization standard, making it highly effective for complex configuration tasks. The input file contains essential execution information, such as the random seed value, the output file path, and the chosen fitting method. It also dictates the selection of the fitting module and the specific PARTONS modules, which govern, in particular, the evaluation of $\xi$ and $\mu^2$, as well as fully differential cross-sections, thereby establishing the complete computational chain. Additionally, the file specifies the choice of minimizer and the initial parameter values, among other settings.

The extraction framework utilizes data retrieved from the Open Database for Exclusive Physics~\cite{Burkert:2025gzu}. This database constitutes another open-source initiative within the broader computational environment illustrated in Fig.~\ref{fig:extraction:workflow}. Because the database is structured around predefined data types, operations such as associating specific processes and observables with their corresponding PARTONS modules, performing unit conversions, and propagating uncertainties are significantly streamlined. Data selection, encompassing tasks such as the isolation of specific datasets and functional dependencies, the execution of kinematic cuts, and the rejection of outliers, is entirely governed by the extraction framework's input YAML file. Although the database is implemented in Python, it interfaces seamlessly with the extraction framework through a dedicated C++ binding.

The extraction framework generates output in the form of an HDF5 file, a standard hierarchical format optimized for complex scientific data storage. To guarantee strict computational reproducibility, this file embeds the original YAML configuration alongside the obtained results. Specifically, it records the optimized values and associated uncertainties of all constrained parameters, exact references to the experimental data utilized during the fitting procedure, and the observables computed by the constrained model. This comprehensive data structure significantly streamlines subsequent postprocessing analyses, which can be executed entirely in Python, an environment that also readily interfaces with the aforementioned database. For instance, via a concise script, one can easily analyze pull distributions, evaluate $\chi^2$ values per dataset, and generate basic plots comparing the fitted data to the corresponding model results.

The output HDF5 files generated by the framework are directly readable by PARTONS, providing an optimal format for disseminating the obtained results. Since PARTONS serves as the foundation for both the extraction framework and the EpIC Monte Carlo generator~\cite{Aschenauer:2022aeb} (another open-source project), the constrained CFF models developed in this work can be subsequently utilized by experimental groups. This includes the evaluation of acceptance corrections and precise studies of future measurements, particularly those planned for next-generation facilities.

\section{Fitting Ansätze and methods}
\label{sec:ansatz}

\subsection{Theory-agnostic model}
\label{sec:ansatz:model_non_par}

The theory-agnostic model is constructed without any physical assumptions. It is utilized to independently extract the real and imaginary parts of the CFFs, relying solely on the statistical power of the experimental data. Such an agnostic method enables an unbiased extraction and was previously employed in Ref.~\cite{Moutarde:2019tqa} on a subset of the data analyzed in the current study.

The Ansätze for the real and imaginary parts of a generic CFF, $\mathcal{F}$, are parameterized by two independent ANNs, each featuring a single hidden layer composed of $n$ neurons. The input to each network is a 6-dimensional vector containing the scaled kinematic variables and their logarithmic transformations: 
\begin{equation}
    \vec{x}_{\mathrm{6D}} = (\overline\xi, \overline{\log_{10}\xi}, \overline{|t|}, \overline{\log_{10}|t|}, \overline{Q^2}, \overline{\log_{10}Q^2})^T\,.
\end{equation}
The scaling is introduced to improve training stability and performance. Specifically, the components are linearly scaled to approximately the range $(-1, 1)$ before being fed into the network: 
\begin{equation}
    \overline{v} = -1 + 2 \frac{v - v_{\mathrm{min}}}{v_{\mathrm{max}} - v_{\mathrm{min}}}\,,
\end{equation}
where $v \in \{\xi, \log_{10}\xi, |t|, \log_{10}|t|, Q^2, \log_{10}Q^2\}$. The values of $v_{\mathrm{min}}$ and $v_{\mathrm{max}}$ are summarized in Table~\ref{tab:kinematic_ranges}.
\begin{table}[!ht]
    \centering
    \caption{Minimum and maximum values of the kinematic variables and their logarithmic transformations used for network input scaling.}
    \label{tab:kinematic_ranges}
    \begin{tabular}{lcc @{\qquad} lcc @{\qquad} lcc}
        \toprule
        $v$ & $v_{\mathrm{min}}$ & $v_{\mathrm{max}}$ & 
        $v$ & $v_{\mathrm{min}}$ & $v_{\mathrm{max}}$ & 
        $v$ & $v_{\mathrm{min}}$ & $v_{\mathrm{max}}$ \\
        \midrule
        $\xi$ & $0$ & $1$ & 
        $t / \mathrm{GeV}^2$ & $-1$ & $0$ &
        $Q^2 / \mathrm{GeV}^2$ & $1$ & $10$ \\
        $\log_{10}\xi$ & $-6$ & $0$ & 
        $\log_{10}(-t / \mathrm{GeV}^2)$ & $-6$ & $0$ &
        $\log_{10}(Q^2 / \mathrm{GeV}^2)$ & $0$ & $1$ \\
        \bottomrule
    \end{tabular}
\end{table}

We utilize both linear kinematic variables and their logarithmic counterparts to accurately capture the dynamics of CFFs across both small and large values of $\xi$, $t$, and $Q^2$. For instance, capturing the small-$\xi$ behavior ($\xi < 0.01$) would require an overly complex and flexible network if the inputs depended only linearly on this variable. Conversely, relying solely on a logarithmic transformation would present a similar problem in describing the large-$\xi$ behavior ($\xi > 0.5$).

\begin{figure}[!ht]
    \centering
    \includegraphics[width=0.5\textwidth]{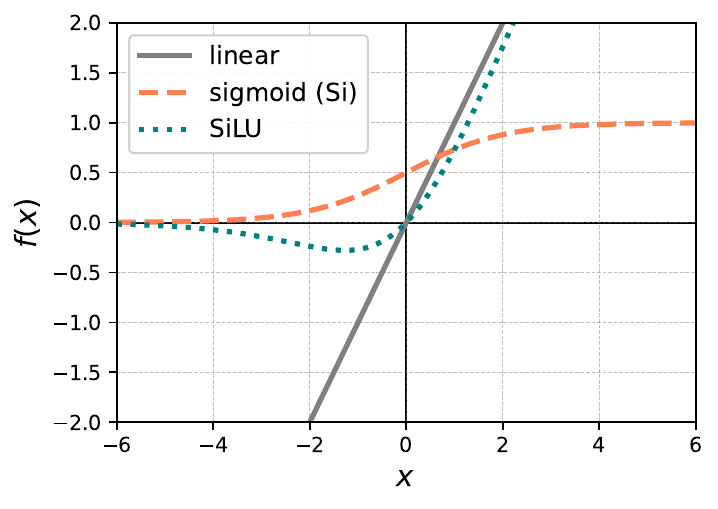}
    \caption{Various types of activation functions used in this analysis.}
    \label{fig:ansatz:activation_function}
\end{figure}

In matrix notation, the pre-scaled outputs of the networks are given by:
\begin{align}
    \overline{O}^{(\mathrm{Re})}(\xi, t, Q^2) &= \frac{1}{n} {W}^{(\mathrm{Re})}_{\mathrm{out}} \mathrm{SiLU}\left({W}^{(\mathrm{Re})}_{\mathrm{in}} \vec{x}_{\mathrm{6D}} + \vec{b}^{(\mathrm{Re})}_{\mathrm{in}} \right) + b^{(\mathrm{Re})}_{\mathrm{out}} \,, \\
    \overline{O}^{(\mathrm{Im})}(\xi, t, Q^2) &= \frac{1}{n} {W}^{(\mathrm{Im})}_{\mathrm{out}} \mathrm{SiLU}\left({W}^{(\mathrm{Im})}_{\mathrm{in}} \vec{x}_{\mathrm{6D}} + \vec{b}^{(\mathrm{Im})}_{\mathrm{in}} \right) + b^{(\mathrm{Im})}_{\mathrm{out}} \,.
\end{align}
The matrices ${W}_{\mathrm{in}} \in \mathbb{R}^{n \times 6}$ and ${W}_{\mathrm{out}} \in \mathbb{R}^{1 \times n}$ denote the weights of the hidden (inner) and output layers, respectively, while $\vec{b}_{\mathrm{in}} \in \mathbb{R}^n$ and $b_{\mathrm{out}} \in \mathbb{R}$ denote the corresponding biases. The activation function applied element-wise within the hidden layer is the sigmoid linear unit (SiLU), defined as:
\begin{equation}
    \mathrm{SiLU}(x) = \frac{x}{1 + e^{-x}}\,,
\end{equation}
which is depicted in Fig.~\ref{fig:ansatz:activation_function}.

A linear scaling is applied to the final output of the network. Assuming the normalized network yields a value roughly within $(-1, 1)$, it is mapped back to the physical output domain of the CFF multiplied by $\xi$, $(O_{\mathrm{min}}, O_{\mathrm{max}}) = (-5, 5)$:
\begin{align}
    \xi\mathrm{Re}\mathcal{F}(\xi, t, Q^2) &= O_{\mathrm{min}} + \frac{1}{2} (\overline{O}^{(\mathrm{Re})}(\xi, t, Q^2) + 1)(O_{\mathrm{max}} - O_{\mathrm{min}})\,, 
    \label{eq:ansatz:scaling_back_re} \\
    \xi\mathrm{Im}\mathcal{F}(\xi, t, Q^2) &= O_{\mathrm{min}} + \frac{1}{2} (\overline{O}^{(\mathrm{Im})}(\xi, t, Q^2) + 1)(O_{\mathrm{max}} - O_{\mathrm{min}})\,.
    \label{eq:ansatz:scaling_back_im}
\end{align}

The free parameters of the Ansatz, namely, the weights and biases, are optimized during the training phase, as detailed in Sect.~\ref{sec:ansatz:minimisation}. Prior to training, the weights are randomly initialized from a normal 
distribution $\mathcal{N}(\mu=0, \sigma=2.5)$ to break the initial  symmetry between the nodes. Simultaneously, the biases are initialized to zero to ensure the initial pre-activations remain zero-centered, preventing premature saturation. Together, these choices enable stable and effective gradient-based optimization.

\subsection{Theory-augmented model}
\label{sec:ansatz:model_par}

In the case of the theory-augmented model, we impose the following theory-driven constraints:
\begin{itemize}
    \item dispersion relations: exploiting dispersion relations eliminates the need to independently parameterize the real and imaginary parts of the CFFs $\mathcal{H}^{++}$, $\mathcal{E}^{++}$, $\widetilde{\mathcal{H}}^{++}$, and $\widetilde{\mathcal{E}}^{++}$ (which would otherwise require eight 3D functions). Instead, it suffices to model the imaginary parts and a single subtraction constant, reducing the problem to four 3D functions and one 2D function. Because the dispersion relations for the remaining two CFFs ($\mathcal{H}^{+-}$ and $\mathcal{H}^{0+}$) are currently unknown, we constrain their real and imaginary parts separately.
\end{itemize}
Furthermore, we impose several additional constraints guided by the standard modeling of GPDs, their connections to PDFs and EFFs, and their ultimate impact on CFFs:
\begin{itemize}
    \item the imaginary parts of the CFFs receive distinct contributions from valence quarks, sea quarks, and gluons. These contributions are characterized by the power-law form $\xi^\alpha (1-\xi)^\beta$, which vanishes at $\xi=1$ and may exhibit a singularity as $\xi \to 0$. The exponent $\alpha$ differs significantly between the valence and sea/gluon sectors, yielding distinct $\xi$-dependent profiles for each contribution to the CFFs.   
    \item kinematic dependencies on both $Q^2$ and $t$ are introduced into the exponent $\alpha$, reflecting the evolution of the PDFs and the Regge behaviour, respectively.
    \item an dipole-like falloff with respect to $-t$ is enforced, mirroring the well-established relationship between GPDs and EFFs for the nucleon.
\end{itemize}

These physical considerations motivate the following Ansatz for the pre-scaled imaginary parts of the CFFs:
\begin{align}
    \overline{O}^{(\mathrm{Im})}(\xi, t, Q^2) &= \left( \sum_{k=1}^{n_{\mathrm{net}}} \xi^{\alpha_k(t, Q^2)} (1-\xi)^{\beta_k} \mathrm{ANN}^{(\mathrm{Im}, k)}(\vec{x}_{2\mathrm{D}}) \right) \nonumber \\
    & \times \left( 1 - \frac{t}{\Lambda_{\mathrm{Im}}^2}\right)^{-\mathrm{ANN}^{(\mathrm{Im}, t)}(\vec{x}_{2\mathrm{D}})}\,,
\end{align}
where $\Lambda_{\mathrm{Im}}$ is fitted to the data, and
\begin{align}
    \alpha_k(t, Q^2) &= \alpha_{k, 0} + \alpha_{k, t} t + \alpha_{k, Q^2} \ln(Q^2/Q^2_0) \,,
\end{align}
with $Q^2_0 = 2\,\mathrm{GeV}^2$. 

We employ $n_{\mathrm{net}} = 2$ independent networks to describe the residual shape of the CFFs beyond the assumed power-law forms, which roughly correspond to the valence and sea/gluon sectors. Our Ansatz is scalable, allowing a higher value for $n_{\mathrm{net}}$ to be adopted if required by the data.

The inputs to the networks are two-dimensional vectors, 
\begin{equation}
    \vec{x}_{2\mathrm{D}} = (\overline{\log_{10}\xi}, \overline{\log_{10}Q^2})^T \,,
\end{equation}
whose elements rely on the previously defined scaling transformations using the parameters summarized in Table~\ref{tab:kinematic_ranges}. Unlike in the theory-agnostic model, we do not use both linear and logarithmic variables. This is because the present model is inherently constrained at the endpoints. In particular, the imaginary parts of the CFFs vanish identically as $\xi \to 1$.

In matrix notation, the ANNs are defined by the following equations:
\begin{align}
    \mathrm{ANN}^{(\mathrm{Im}, k)}(\vec{x}_{\mathrm{2D}}) &= \frac{1}{n} W^{(\mathrm{Im}, k)}_{\mathrm{out}} \mathrm{SiLU}\left(W^{(\mathrm{Im}, k)}_{\mathrm{in}} \vec{x}_{\mathrm{2D}} + \vec{b}^{(\mathrm{Im}, k)}_{\mathrm{in}} \right) + b^{(\mathrm{Im}, k)}_{\mathrm{out}}\,, \\
    \mathrm{ANN}^{(\mathrm{Im}, t)}(\vec{x}_{\mathrm{2D}}) &= \frac{1}{n} W^{(\mathrm{Im}, t)}_{\mathrm{out}} \mathrm{Si}\left(W^{(\mathrm{Im}, t)}_{\mathrm{in}} \vec{x}_{\mathrm{2D}} + \vec{b}^{(\mathrm{Im}, t)}_{\mathrm{in}} \right) + b^{(\mathrm{Im}, t)}_{\mathrm{out}}\,.
\end{align}
Here, for simplicity, we assume that each network consists of the same number of hidden neurons, $n$. For the shape networks ($k \in \{1, \dots, n_{\mathrm{net}}\}$), the matrices $W^{(\mathrm{Im}, k)}_{\mathrm{in}} \in \mathbb{R}^{n \times 2}$ and $W^{(\mathrm{Im}, k)}_{\mathrm{out}} \in \mathbb{R}^{1 \times n}$ denote the weights of the hidden and output layers, respectively, with $\vec{b}^{(\mathrm{Im}, k)}_{\mathrm{in}} \in \mathbb{R}^{n}$ and $b^{(\mathrm{Im}, k)}_{\mathrm{out}} \in \mathbb{R}$ representing the corresponding biases. Similarly, the parameters for the network related to the $t$-dependence are given by $W^{(\mathrm{Im}, t)}_{\mathrm{in}} \in \mathbb{R}^{n \times 2}$, $W^{(\mathrm{Im}, t)}_{\mathrm{out}} \in \mathbb{R}^{1 \times n}$, $\vec{b}^{(\mathrm{Im}, t)}_{\mathrm{in}} \in \mathbb{R}^{n}$, and $b^{(\mathrm{Im}, t)}_{\mathrm{out}} \in \mathbb{R}$.

While the shape networks utilize the SiLU activation function, the network used to describe the $t$-dependence employs a sigmoid activation function:
\begin{equation}
    \mathrm{Si}(x) = \frac{1}{1 + e^{-x}} \,.
\end{equation}
This choice provides a more stable training environment by naturally bounding the output to the $(0, 1)$ range (see Fig.~\ref{fig:ansatz:activation_function}). Constraining the outer biases $b^{(\mathrm{Im}, t)}_{\mathrm{out}}$ and weights $W^{(\mathrm{Im}, t)}_{\mathrm{out}}$ to be strictly positive guarantees that the Ansatz does not diverge as $-t \to \infty$. The linear output scaling used to map $\overline{O}^{(\mathrm{Im})}$ back to the physical domain remains identical to that of the theory-agnostic model, see Eq.~\eqref{eq:ansatz:scaling_back_im}.

The real parts of the CFFs $\mathcal{H}^{+-}$ and $\mathcal{H}^{0+}$ are parameterized exactly like the imaginary parts, with the exception of setting the exponents $\beta_k = 0$. This removes the requirement for the functions to vanish at $\xi = 1$. Unlike the imaginary parts, which are kinematically restricted to vanish at this boundary in the direct channel, the real parts are not bound by this restriction due to their analytic continuation into crossed channels (see for instance Ref.~\cite{Semenov-Tian-Shansky:2023ysr}).

Finally, the subtraction constant, which depends exclusively on $t$ and $Q^2$, is parameterized as the product of a $Q^2$-dependent parameter describing the magnitude at $t=0$ and, as previously, a dipole-like structure describing the falloff in $-t$:
\begin{equation}
    \mathcal{S}(t, Q^2) = \mathcal{S}(Q^2) \left( 1 - \frac{t}{\Lambda_{\mathrm{SC}}^2}\right)^{-\mathrm{ANN}^{(\mathrm{SC}, t)}(x_{\mathrm{1D}})} \,.
\end{equation}
The neural network $\mathrm{ANN}^{(\mathrm{SC}, t)}(x_{\mathrm{1D}})$ takes only the scaled logarithmic $Q^2$ variable as input,
\begin{equation}
    x_{\mathrm{1D}} \equiv \overline{\log_{10}Q^2} \,,
\end{equation}
and in matrix notation it is defined by the following equation:
\begin{equation}
    \mathrm{ANN}^{(\mathrm{SC}, t)}(x_{\mathrm{1D}}) = \frac{1}{n} W^{(\mathrm{SC}, t)}_{\mathrm{out}} \mathrm{Si} \left(W^{(\mathrm{SC}, t)}_{\mathrm{in}} x_{\mathrm{1D}} + \vec{b}^{(\mathrm{SC}, t)}_{\mathrm{in}} \right) + b^{(\mathrm{SC}, t)}_{\mathrm{out}}\,.
\end{equation}
Here, $n$ again represents the number of hidden neurons. The matrices $W^{(\mathrm{SC}, t)}_{\mathrm{in}} \in \mathbb{R}^{n \times 1}$ and $W^{(\mathrm{SC}, t)}_{\mathrm{out}} \in \mathbb{R}^{1 \times n}$ denote the weights of the hidden and output layers, respectively, with $\vec{b}^{(\mathrm{SC}, t)}_{\mathrm{in}} \in \mathbb{R}^{n}$ and $b^{(\mathrm{SC}, t)}_{\mathrm{out}} \in \mathbb{R}$ representing the corresponding biases.

The magnitude at $t=0$ explicitly mimics the $Q^2$ evolution and is defined as:
\begin{equation}
    \mathcal{S}(Q^2) = \mathcal{S}_{0} + \frac{1}{n} \sum_{i=1}^{n} \mathcal{S}_i \left( \frac{\ln(Q^2/\Lambda_{\mathrm{QCD}}^2)}{\ln(Q_0^2/\Lambda_{\mathrm{QCD}}^2)} \right)^{p_i} \,,
    \label{eq:ansatz:model_par:sc}
\end{equation}
where $\Lambda_{\mathrm{QCD}} = 0.2\,\mathrm{GeV}$ is the QCD scale parameter, $Q_0^2 = 2\,\mathrm{GeV}^2$ is the reference scale, and $p_i$ are free parameters controlling the effective $Q^2$ evolution.

Prior to the optimization phase, all free parameters of the theory-augmented Ansatz are initialized using specific statistical distributions. The baseline powers governing the dominant power-law behaviour are initialized from uniform distributions, with $\alpha_{1,0} \sim \mathcal{U}(a=0, b=0.5)$ for the first network term, $\alpha_{2,0} \sim \mathcal{U}(-0.5, 0)$ for the second, and $\beta_{k} \sim \mathcal{N}(0.5, 0.1)$ for both $k \in \{1, 2\}$. Determining appropriate initial values for $\beta_{k}$ poses a challenge because excessively high values force the distributions to vanish too rapidly as $\xi \to 1$, affecting the uncertainty estimation in that region. The remaining parameters dictating the kinematic evolution of the baseline powers ($\alpha_{k,t}$ and $\alpha_{k,Q^2}$) are drawn from a narrow normal distribution, $\mathcal{N}(0, 0.1)$, to introduce small initial perturbations centered around zero. The mass scale parameter $\Lambda_{\mathrm{Im}}^2$ is initialized from $\mathcal{N}(1, 0.1)$.

For the neural networks describing the residual shape of the CFFs ($\mathrm{ANN}^{(\mathrm{Im}, k)}$ with $k \in \{1, 2\}$), the weights $W^{(\mathrm{Im}, k)}_{\mathrm{in}}$ and $W^{(\mathrm{Im}, k)}_{\mathrm{out}}$ are drawn from $\mathcal{N}(0, 0.5)$, while their global output biases are initialized from a standard normal distribution as $b^{(\mathrm{Im}, k)}_{\mathrm{out}} \sim \mathcal{N}(0, 1)$. To model the global $t$-dependence and mimic standard dipole/tripole behaviour, the weights $W^{(\mathrm{Im}, t)}_{\mathrm{in}}$ and $W^{(\mathrm{Im}, t)}_{\mathrm{out}}$ of $\mathrm{ANN}^{(\mathrm{Im}, t)}$ are initialized from $\mathcal{N}(0, 0.5)$, with its global output bias drawn as $b^{(\mathrm{Im}, t)}_{\mathrm{out}} \sim \mathcal{N}(3, 0.1)$. To break symmetry without introducing unnecessary early saturation, the inner biases ($\vec{b}^{(\mathrm{Im}, k)}_{\mathrm{in}}$ and $\vec{b}^{(\mathrm{Im}, t)}_{\mathrm{in}}$) for the hidden neurons in all three networks are strictly initialized to zero.

Finally, the coefficients defining the subtraction constant are initialized according to distinct distributions to reflect their physical roles. The constant amplitude parameter $\mathcal{S}_{0}$ and the amplitude weights $\mathcal{S}_i$ are drawn from a uniform distribution, $\mathcal{U}(-10, 10)$. To strictly ensure an asymptotically vanishing behaviour at high $Q^2$, 
the parameters $p_i$ are forced to be negative and are initialized 
from a uniform distribution $\mathcal{U}(-5, -1)$. The subtraction constant's mass scale parameter $\Lambda_{\mathrm{SC}}^2$ is drawn from $\mathcal{N}(1, 0.1)$. Analogous to the imaginary part, the $t$-dependence network $\mathrm{ANN}^{(\mathrm{SC}, t)}$ for the subtraction constant has its weights $W^{(\mathrm{SC}, t)}_{\mathrm{in}}$ and $W^{(\mathrm{SC}, t)}_{\mathrm{out}}$ initialized from $\mathcal{N}(0, 0.5)$, its inner biases $\vec{b}^{(\mathrm{SC}, t)}_{\mathrm{in}}$ strictly set to zero, and its global output bias $b^{(\mathrm{SC}, t)}_{\mathrm{out}}$ initialized from $\mathcal{N}(3, 0.1)$.

\subsection{Propagation of uncertainties}
\label{sec:ansatz:uncertainties}

Uncertainties are propagated from the experimental data using the standard replication method. Specifically, we perform the fit $101$ times. The first fit, referred to as ``the central replica'', is performed on 
the original, unmodified data to obtain the central values of the model 
parameters. The remaining $100$ fits (replicas) are then performed on 
artificially fluctuated pseudodata. For the $i$-th replica, the value of the $j$-th point belonging to dataset $k$ is generated according to:
\begin{equation}
    v^{(i)}_{j/k} = v_{j/k} + \mathcal{N}^{(i)}_{j/k}(0, 1)\sqrt{\sigma_{\mathrm{stat}, j/k}^2 + \sigma_{\mathrm{sys}, j/k}^2} + \mathcal{N}^{(i)}_{k}(0, 1)\sigma_{\mathrm{norm}, j/k} \,.
\end{equation}
Here, $v_{j/k}$, $\sigma_{\mathrm{stat}, j/k}$, $\sigma_{\mathrm{sys}, j/k}$, and $\sigma_{\mathrm{norm}, j/k}$ represent the central value, statistical uncertainty, systematic uncertainty, and normalization uncertainty of a given experimental point, respectively. The normalization uncertainty is typically associated with luminosity or polarization measurements. 

The terms $\smash{\mathcal{N}^{(i)}_{j/k}(0, 1)}$ and $\smash{\mathcal{N}^{(i)}_{k}(0, 1)}$ denote random numbers drawn from a standard normal distribution. Crucially, $\smash{\mathcal{N}^{(i)}_{j/k}}$ is sampled independently for every single point, while $\smash{\mathcal{N}^{(i)}_{k}}$ is sampled only once per dataset for a given replica. This ensures that all points within a specific dataset shift coherently in response to the normalization uncertainty, correctly capturing its fully correlated nature. We treat the remaining systematic uncertainties as uncorrelated point-to-point, as the experimental publications generally do not provide sufficient covariance information to model them otherwise.  Consequently, the total uncertainty associated with a given point, which enters the evaluation of $\chi^{2}_{\mathrm{data}}(\theta_i)$ (see Eq.~\eqref{eq:ansatz:chi2_exp}), is given by:
\begin{equation}
    \sigma_{j/k}^{(\mathrm{exp})} = \sqrt{\sigma_{\mathrm{stat}, j/k}^2 + \sigma_{\mathrm{sys}, j/k}^2 + \sigma_{\mathrm{norm}, j/k}^2}\,,
\end{equation}
where we explicitly expose the index $k$ associated with the dataset.

In the presented plots, the uncertainty bands represent the $68\%$ confidence interval. These bands are evaluated from the ensemble of $100$ replicas (excluding the central replica) using the quantile method, specifically by determining the interval bounded by the $16$th and $84$th percentiles.

\subsection{Addressing kinematic sparsity problem}
\label{sec:ansatz:sparsity}

In the global analysis of DVCS, the available experimental data are distributed highly unevenly across the kinematic phase space. Specifically, the low-$\xi$ regime covered by HERA is sparse compared to regions with higher $\xi$ covered by JLAB, which are densely populated by high-precision points. Additionally, the data overwhelmingly populate low $Q^2$ regions compared to high $Q^2$ ones. Standard optimization routines that minimize an unweighted $\chi^2$ loss function inherently prioritize regions with the highest data density. Consequently, the fit may fail to adequately capture the functional behaviour in sparse regions, treating them as statistically insignificant.

To mitigate this and ensure that sparse data, particularly at small $\xi$, remain ``visible'' during the optimization phase, we employ an inverse density weighting procedure. The procedure is executed over a two-dimensional grid spanning the kinematic variables $\xi$ and $Q^2$. Specifically, the phase space is discretized using uniform linear bins for both variables. The variable $\xi$ is partitioned into $200$ bins spanning the range $0 < \xi < 1$, while $Q^2$ is divided into $36$ bins covering the interval $1~\mathrm{GeV}^2 < Q^2 < 10~\mathrm{GeV}^2$. 

To evaluate the weights for the $N_{\mathrm{data}}$ experimental points used in the extraction, we first map them onto the aforementioned grid. Next, we define the ``raw'' weight for the $j$-th data point falling into the $i$-th bin as:
\begin{equation}
    w_{j, \mathrm{raw}} = \frac{1}{N_{\mathrm{unique}, i}} \,,
\end{equation}
where $N_{\mathrm{unique}, i}$ is the number of strictly unique $(\xi, Q^2)$ coordinate pairs registered in the $i$-th bin (note that many experimental points may be characterised by the same values of $\xi$ and $Q^2$). The final, normalized weight $w_j$ assigned to the $j$-th data point is then calculated as:
\begin{equation}
    w_j = N_{\mathrm{data}} \frac{w_{j, \mathrm{raw}}}{\sum_{k} w_{k, \mathrm{raw}}} \,.
\end{equation}
By ensuring that $\sum_{j} w_j = N_{\mathrm{data}}$, the weighting procedure preserves the standard statistical interpretation of the loss function. 

\begin{figure}[!ht]
    \centering
    \includegraphics[width=0.6\textwidth]{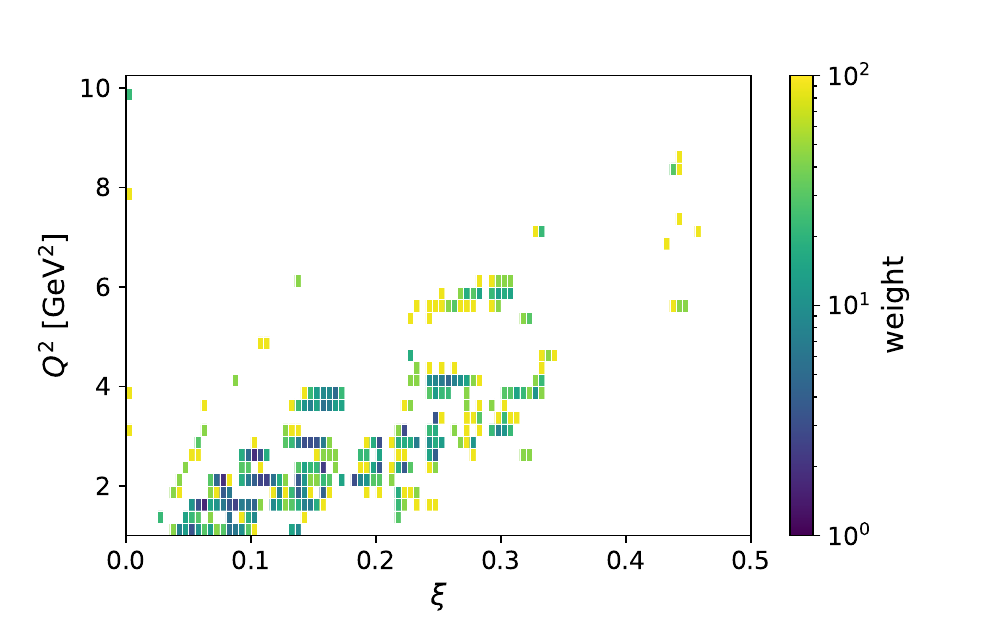}
    \caption{Distribution of weights, $w_j$, across the $(\xi, Q^2)$ kinematic plane.}
    \label{fig:ansatz:plot_weights}
\end{figure}

Note that in the limit of infinitely many bins (or infinitesimal bin sizes), each populated bin isolates exactly one unique kinematic setting, meaning $N_{\mathrm{unique}, i} = 1$ for all populated bins. In this limit, every data point receives a raw weight of $1$, and the procedure naturally and strictly recovers the standard unweighted scenario where $w_j = 1$. This behaviour demonstrates that by changing the binning resolution, we can seamlessly control the strength of the $(\xi, Q^2)$ density reweighting. The distribution of weights used in the current analysis is shown in Fig.~\ref{fig:ansatz:plot_weights}.

During the minimization procedure, the weights are incorporated into the goodness-of-fit function as follows:
\begin{equation}
    \chi^{2}_{\mathrm{data}}(\theta_i) = \frac{1}{N_{\mathrm{data}}}\sum_{j=1}^{N_{\mathrm{data}}} 
    w_j 
    \left( \frac{v_j(\theta_i) - v^{(\mathrm{exp})}_j}{\sigma^{(\mathrm{exp})}_j} \right)^2 \, .
    \label{eq:ansatz:chi2_exp}
\end{equation}
Here, $\theta_i$ denotes a complete set of trainable parameters for the $i$-th replica, $v_j(\theta_i)$ denotes the model predictions, while the $N_{\mathrm{data}}$ experimental data points are characterized by their measured values $v^{(\mathrm{exp})}_j$ and associated uncertainties $\sigma^{(\mathrm{exp})}_j$. In the following, we will also use the unweighted estimator, simply defined as:
\begin{equation}
    \chi^{2}_{\mathrm{data, uw}}(\theta_i) = \frac{1}{N_{\mathrm{data}}}\sum_{j=1}^{N_{\mathrm{data}}} 
    \left( \frac{v_j(\theta_i) - v^{(\mathrm{exp})}_j}{\sigma^{(\mathrm{exp})}_j} \right)^2 \, .
    \label{eq:ansatz:chi2_exp_unweighted}
\end{equation}

\begin{figure}[!ht]
    \centering
    \includegraphics[width=0.49\textwidth]{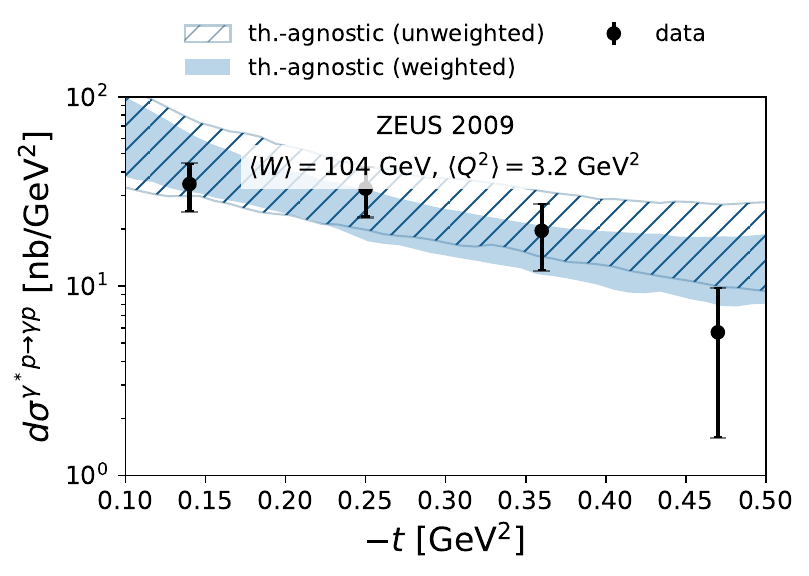}
    \includegraphics[width=0.49\textwidth]{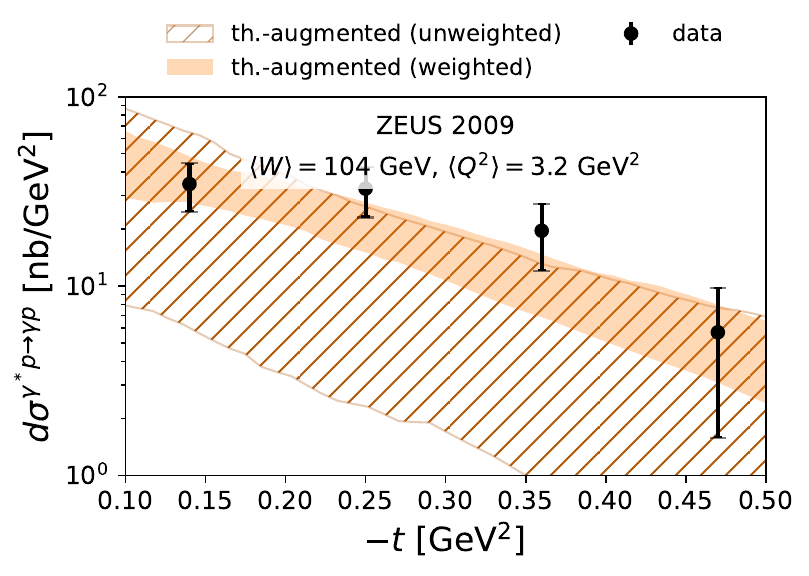}
    \caption{Comparison of results obtained with and without the weighting procedure for the theory-agnostic (left) and theory-augmented (right) models. In both cases, the results are compared to the ZEUS data~\cite{ZEUS:2008hcd}.}
    \label{fig:ansatz:plot_weights_effect}
\end{figure}

The weighting procedure introduces an increase in $\chi^{2}_{\mathrm{data}, \mathrm{uw}}$ of approximately $1\%$ compared to a fully unweighted extraction. The effect of this procedure is demonstrated in Fig.~\ref{fig:ansatz:plot_weights_effect}. 
For the theory-agnostic model, the impact is moderate, as expected. We primarily 
observe a reduction in the artificially inflated uncertainties associated with 
the unweighted extraction, an issue that could otherwise only be mitigated by 
using an unrealistically long training cycle and large networks. For the theory-augmented model, 
however, the effect is striking: the weighting procedure successfully prevents 
both bias and the inflation of uncertainties. It is also worth noting that the results for both models are in better agreement when the weighting procedure is used.

\subsection{Addressing bias-variance problem}
\label{sec:ansatz:bias_variance}

A fundamental challenge in employing artificial neural networks is managing the bias-variance trade-off. Due to their immense structural flexibility, ANNs inherently exhibit low bias but suffer from high variance, making them highly susceptible to overfitting the experimental data and capturing statistical noise. While traditional regularization techniques, such as standard L1 and L2 weight decay, effectively reduce variance by penalizing the magnitude of the network weights, they indiscriminately pull all parameters toward zero, resulting in underestimated and therefore unreliable uncertainties in domains not covered by data.

To systematically address this, we employ a technique known as anchored ensembling. Instead of penalizing the absolute magnitude of the trainable weights, we penalize their deviation from a set of fixed, randomly initialized anchor values. For each replica, we instantiate a set of corresponding prior networks alongside the trainable networks, keeping the parameters of these prior networks frozen. The total loss function optimized during the training of the $i$-th replica is thus modified to include an anchoring penalty:
\begin{align}
    \chi^{2}(\theta_i) = \chi^{2}_{\mathrm{data}}(\theta_i) + \lambda\, \chi^2_{\mathrm{anchor}}(\theta_i) \,,
    \label{eq:ansatz:chi2_total}
\end{align}
where the data term, $\chi^{2}_{\mathrm{data}}(\theta_i)$, was defined in Eq.~\eqref{eq:ansatz:chi2_exp}. The anchoring penalty is defined as 
\begin{align}
    \chi^2_{\mathrm{anchor}}(\theta_i) = \frac{1}{N_{\mathrm{par}}}\sum_{j=1}^{N_{\mathrm{par}}} \left(\theta_{i, j} - \theta^{(\mathrm{prior})}_{i, j}\right)^2 \,.
    \label{eq:ansatz:chi2_anchor}
\end{align}
Here, $\theta_i$ denotes the complete set of $N_{\mathrm{par}}$ trainable parameters for the model, $\theta^{(\mathrm{prior})}_{i}$ represents the fixed parameters of the random prior, and $\lambda$ is a hyperparameter that controls the strength of the regularization.

In summary, by assigning a distinct random prior anchor to each replica, the individual networks are encouraged to explore different regions of the loss landscape, ensuring a diverse ensemble. In kinematic regions that are well constrained by experimental data, the data term dominates the optimization and the dependence on the particular anchor becomes weak, leading the replicas to converge toward similar solutions. In sparsely populated regions, on the other hand, the anchoring term prevents the ensemble from collapsing onto a single arbitrary extrapolation, allowing the spread among replicas to reflect the reduced constraining power of the data.

\subsection{Fixing hyperparameters}
\label{sec:ansatz:hyperparameters}

Implementing the presented framework requires a strict, sequential tuning procedure to prevent hyperparameter coupling. To achieve this, we first fix the intrinsic capacity of the models by tuning the number of hidden neurons, ensuring that the networks possess sufficient flexibility to describe the underlying physics. After the network topology is established, we tune the anchoring hyperparameter, $\lambda$, to properly scale the ensemble variance relative to the experimental data.  

Because directly fitting the model to experimental data is computationally expensive, we tune the hyperparameters using a generated set of CFF pseudodata. This pseudodata is constructed through the following procedure:
\begin{enumerate}
    \item for each kinematic point $(\xi, t, Q^2)$ in the experimental dataset, we perform a local extraction (see, e.g., Ref.~\cite{Dupre:2017hfs}) to determine the CFFs and their associated uncertainties.
    \item for these same kinematic points, we compute the corresponding CFF values using the GK model and leading-order (LO) coefficient functions.
    \item we then smear these GK model CFFs according to the uncertainties obtained in the first step.
\end{enumerate}
We utilize the GK model predictions, rather than the locally extracted central values, because local extraction is unreliable for datasets lacking an explicit $\phi$ dependence, such as the HERMES and HERA data. In such cases, estimating the correlations between the extracted CFFs is exceedingly challenging. Furthermore, since many observables are primarily sensitive to the square of the CFFs, the signs of the locally extracted CFFs remain largely unconstrained and can be entirely random.

We begin by determining the optimal number of hidden neurons. At this stage, we set the anchoring hyperparameter to $\lambda=0$, thereby disabling the influence of the prior distributions. Our goal is to identify the minimum number of neurons required to prevent the model from being too rigid to accurately describe the data. Conversely, we want to avoid excessive flexibility: overly complex models not only become computationally expensive but also require significantly heavier regularization to prevent overfitting, which can introduce unwanted numerical noise.

\begin{figure}[!ht]
    \centering
    \includegraphics[width=0.49\textwidth]{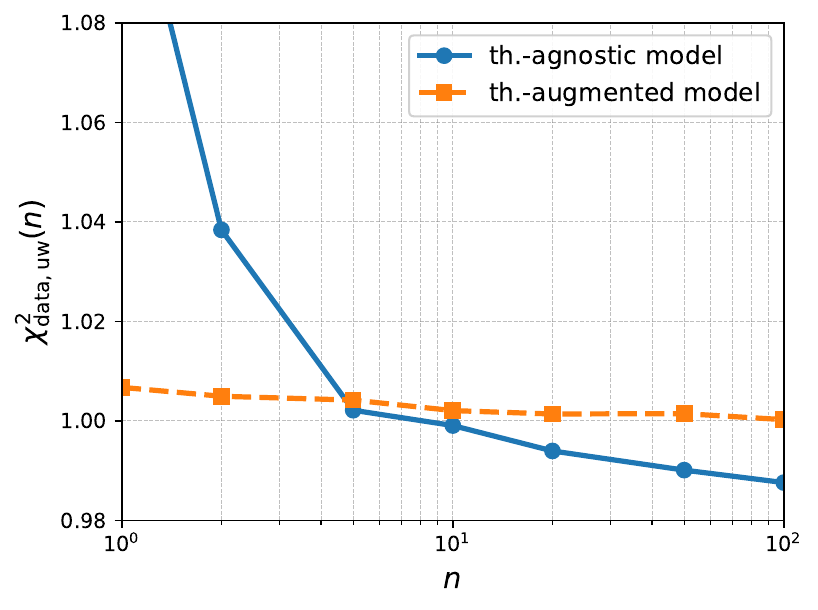}
    \includegraphics[width=0.49\textwidth]{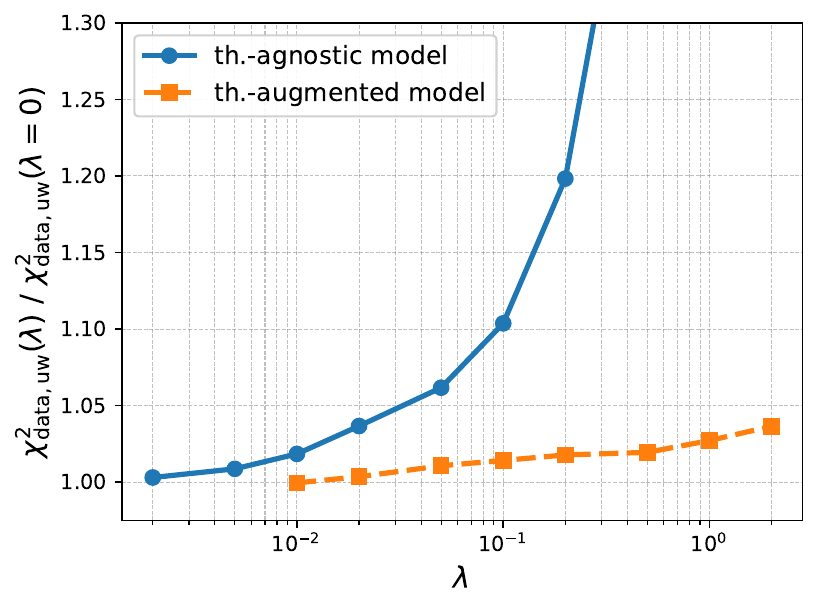}
    \caption{Optimization of hyperparameters. (left) Goodness of fit obtained after minimization (for $\lambda = 0$) as a function of the number of hidden neurons. (right) The same quantity evaluated as a function of the regularization strength $\lambda$, normalized to the unregularized ($\lambda=0$) baseline.}
    \label{fig:ansatz:scan}
\end{figure}

The dependence of the post-training loss $\chi^{2}_{\mathrm{data}, \mathrm{uw}}(\theta_0)$ on the number of neurons is illustrated in Fig.~\ref{fig:ansatz:scan} for both models. For simplicity, in the theory-augmented case, we use the same number of neurons across all constituent networks, as well as an identical number of terms in the subtraction constant ansatz. For instance, $n=5$ indicates five hidden neurons in  $\mathrm{ANN}^{(\mathrm{Im}, k)}(\vec{x}_{\mathrm{2D}})$, where $k=1,2$, and $\mathrm{ANN}^{(\mathrm{Im}, t)}(\vec{x}_{\mathrm{2D}})$, alongside five terms in the sum displayed in Eq.~\eqref{eq:ansatz:model_par:sc}. As shown in the figure, insufficient network capacity restricts the model's ability to accurately describe the data, manifesting as a systematic bias. In the theory-agnostic case, however, highly complex models suffer from overfitting. Consequently, we set $n=10$ for the theory-agnostic model and $n=5$ for the theory-augmented model.

Having determined the optimal network capacities, a process that inherently involves a degree of arbitrariness, we proceed to tune the anchoring hyperparameter $\lambda$. Figure~\ref{fig:ansatz:scan} illustrates the dependence of the post-training loss $\chi^{2}_{\mathrm{data}, \mathrm{uw}}(\theta_0)$ on $\lambda$ for both architectures. As expected, if $\lambda$ is set too large, the optimization is dominated by the regularization penalty, driving the model too strongly toward the random prior and reintroducing a systematic bias. To balance this trade-off, ensuring ensemble diversity without sacrificing fit quality, we set $\lambda = 0.01$ for the theory-agnostic model and $\lambda = 0.1$ for the theory-augmented model.

\subsection{Minimisation procedure}
\label{sec:ansatz:minimisation}

The minimisation utilizes the Adam algorithm~\cite{kingma:2017}. To mitigate the computational and temporal costs of the minimization process, we accelerate convergence using a three-stage procedure (a technique known as warm-starting):
\begin{enumerate}
\item first, we initialize a fully unconstrained model. The weights, biases, and any additional parameters (such as powers in the theory-augmented model) are randomly selected, as detailed in Secs.~\ref{sec:ansatz:model_non_par} and~\ref{sec:ansatz:model_par} introducing the models.
\item second, using this unconstrained model, we perform a preliminary fit to pseudodata for the CFFs based on the GK model (these are the same pseudodata used to fix the hyperparameters, see Sect.~\ref{sec:ansatz:bias_variance}). The rapid fit is terminated once $\chi^{2}_{\mathrm{data}}(\theta_0) < 25$ (with the very last iteration being discarded), ensuring that our initial ensemble of replicas forms a broad band around the CFFs predicted by the GK model.
\item finally, using the parameters obtained from this intermediate step, we resume the fitting process, this time utilizing the actual experimental data.
\end{enumerate}

\begin{figure}[!ht]
    \centering
    \includegraphics[width=0.49\textwidth]{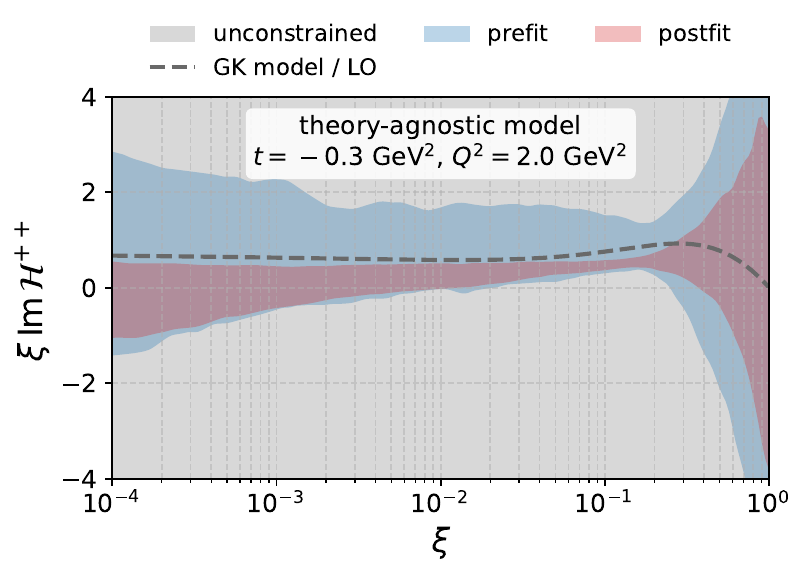}
    \includegraphics[width=0.49\textwidth]{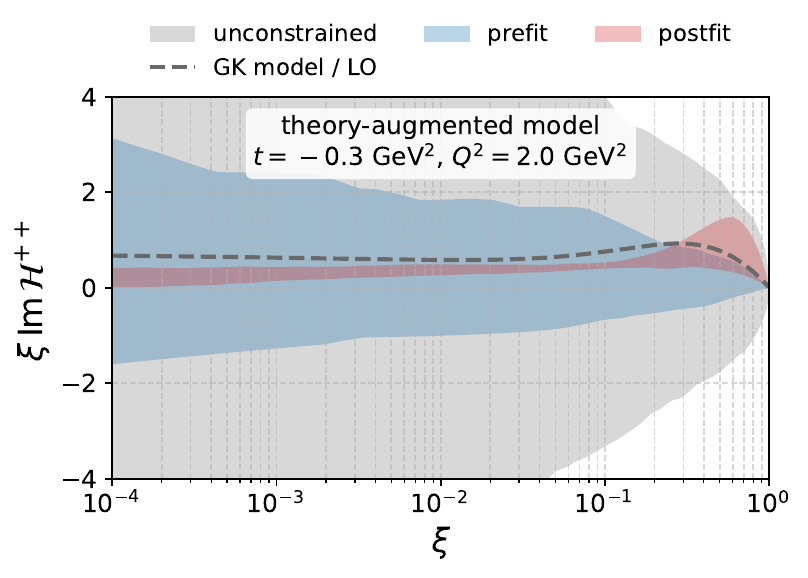}
    \caption{Comparison of the results for the imaginary part of the CFF $\mathcal{H}^{++}$ obtained following the three-stage minimization procedure described in the text.}
    \label{fig:ansatz:fit_stages}
\end{figure}

This procedure is illustrated in Fig.~\ref{fig:ansatz:fit_stages}, which displays the results for the fully unconstrained model, the model constrained by the preliminary fit to pseudo-data, and the final model constrained by the experimental data. Had we fitted the experimental data directly from a fully unconstrained model (i.e., omitting the intermediate step), the initial goodness-of-fit value would be on the order of $\chi^{2}_{\mathrm{data}}(\theta_0)  \sim 10^{7}$, significantly prolonging convergence. Instead, our intermediate step reduces the starting point for the final fit to approximately $\chi^{2}_{\mathrm{data}}(\theta_0) \sim 10^{3}$. 
Only for the theory-augmented model, the prefit procedure encounters difficulties at large $\xi$ due to a strong non-linear dependence of CFFs on the powers $\beta_k$ in the term $(1-x)^{\beta_k}$. Specifically, to ensure a broad uncertainty band as $\xi \to 1$, the values of $\beta_k$ must be small. Although they are initialized appropriately, the gradient-based training algorithm used in the prefit tends to increase these parameters on average due to the scarcity of data at large $\xi$. This artificial behaviour, however, does not impact the final extraction.

\begin{figure}[!ht]
    \centering
    \includegraphics[width=0.49\textwidth]{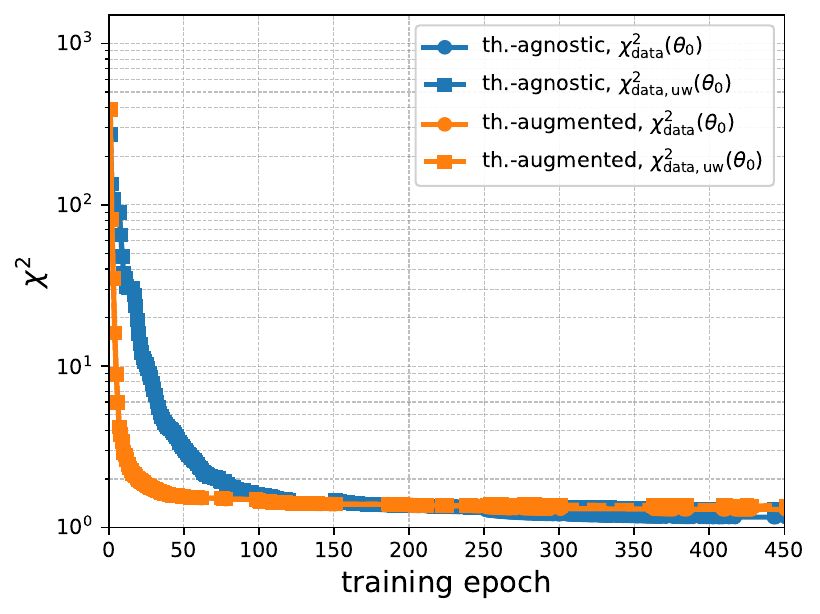}
    \includegraphics[width=0.49\textwidth]{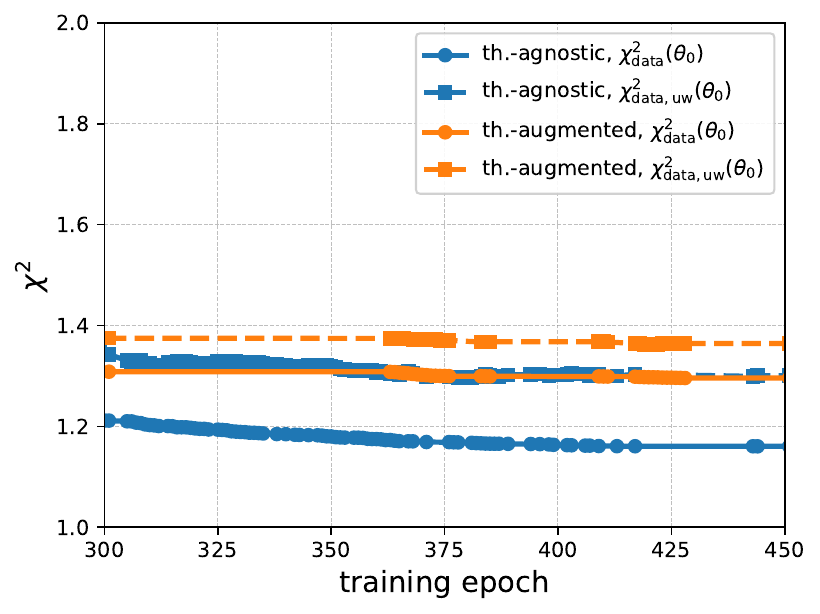}
    \caption{Goodness of fit as a function of the training epoch. (left) The full training range. (right) A magnified view of the left panel.}
    \label{fig:ansatz:plot_t}
\end{figure}

The minimization algorithm employs a batching strategy and is initialized with a learning rate of $0.05$. To dynamically refine the optimization, this rate is halved if the fit fails to improve for $20$ consecutive iterations. Additionally, we set the optimizer hyperparameters to $\beta_1 = 0.9$ and $\beta_2 = 0.9$, which dictate the exponential decay rates for the first and second moment estimates (tracking the moving averages of the gradient and the squared gradient, respectively). This specific configuration ensures stable and rapid convergence. The typical evolution of the losses $\chi^{2}_{\mathrm{data}}(\theta_0)$ and $\chi^{2}_{\mathrm{data}, \mathrm{uw}}(\theta_0)$ as a function of the training epoch during the fit to experimental data is shown in Fig.~\ref{fig:ansatz:plot_t} for both models.

\section{Result}
\label{sec:result}

\subsection{Selection of data and fit performance}

In this analysis, we extract DVCS CFFs from experimental data separately for proton and neutron targets. A meaningful joint extraction for both targets can only be performed at the GPD level, unless one employs a more involved parameterization of CFFs that incorporates flavour separation, an approach inherently more susceptible to model dependencies. For a similar reason, we do not include the existing TCS data~\cite{CLAS:2021lky} in the current fit. Including these data would necessitate adopting specific relations between the DVCS and TCS amplitudes, thereby forcing us to perform the extraction at a fixed perturbative order in $\alpha_{s}$. Nevertheless, a comparison with the existing TCS data is presented at the end of this section.

For the fitting procedure, we restrict our analysis to data that satisfy the following kinematic cuts:
\begin{align}
    Q^2 > 1~\mathrm{GeV}^2\,, \quad -t/Q^2 < 0.4\,.
\end{align}
These kinematic criteria are modified with respect to our previous works and those by other phenomenology groups (see, for instance, Refs.~\cite{Kumericki:2015lhb}). Namely, compared to Refs.~\cite{Moutarde:2019tqa}, we relaxed the cuts from $Q^2 > 1.5~\mathrm{GeV}^2$ and $-t/Q^2 < 0.2$. In doing so, we have made the extraction more sensitive to both genuine and kinematic higher-twist effects. This approach is reasonable because, in this analysis, we explicitly include the kinematically suppressed helicity-flip $\mathcal{H}^{0+}$ and $\mathcal{H}^{+-}$ amplitudes, and we aim to match the extraction to the precision of the current understanding of DVCS in the language of GPDs~\cite{Braun:2025xlp}.

The DVCS data considered in this study are summarized in Table~\ref{tab:dvcs_data} for both proton and neutron targets. This table also specifies the number of data points that passed the imposed kinematic cuts, as well as the post-fit $\chi^{2}_{\mathrm{data}, \mathrm{uw}}(\theta_0)$ values obtained for the central replicas of both models. For the proton target, the total $\chi^{2}_{\mathrm{data}, \mathrm{uw}}(\theta_0)$ is $1.29$ for the theory-agnostic model and $1.36$ for the theory-augmented model. The corresponding values for the neutron target are $0.76$ and $0.70$. For datasets common to both the present and our previous analysis~\cite{Moutarde:2019tqa}, the resulting goodness-of-fit values are in good agreement. However, for the newly included data, particularly those from Ref.~\cite{CLAS:2018bgk,CLAS:2022syx}, we generally observe higher $\chi^{2}_{\mathrm{data}, \mathrm{uw}}(\theta_0)$ values. Given that the theory-agnostic model minimises the risk of an inadequate Ansatz, this discrepancy points to either the presence of neglected theoretical contributions or residual systematic issues within the data themselves.
\begin{table}[htbp]
    \centering
    \caption{Summary of the DVCS datasets included in the fit. The datasets are grouped by target type, collaboration, and publication year, and are presented alongside their corresponding post-fit $\chi^{2}_{\mathrm{data}, \mathrm{uw}}(\theta_0)$ values.}
    \label{tab:dvcs_data}
    \renewcommand{\arraystretch}{1.1}
    \begin{tabular}{l c l l c c c}
        \toprule
        \multirow{2}{*}{\textbf{Collab.}} & \multirow{2}{*}{\textbf{Year}} & \multirow{2}{*}{\textbf{Observable}} & \multirow{2}{*}{\textbf{Ref.}} & \multirow{2}{*}{$\mathbf{N_{\mathrm{data}}}$} & \multicolumn{2}{c}{$\mathbf{\chi^{2}_{\mathrm{data, uw}}}(\theta_0)$} \\
        & & & & & \textbf{th.-agnostic} & \textbf{th.-augmented} \\
        \midrule
        \multicolumn{7}{c}{\textbf{PROTON TARGET}} \\
        \midrule
        CLAS 
        & 2001 & $A_{LU}^{\sin\phi}, A_{LU}^{\sin 2\phi}$ & \cite{CLAS:2001wjj} 
                                                        & 2 & 0.80 & 0.62 \\
        & 2006 & $A_{UL}^{\sin\phi}, A_{UL}^{\sin 2\phi}$ & \cite{CLAS:2006krx} 
                                                        & 2 & 0.04 & 2.01 \\
        & 2007 & $A_{LU}$ & \cite{Girod:2007aa} 
                                                        & 546 & 0.88 & 0.84 \\
        & 2008 & $A_{LU}$ & \cite{CLAS:2008ahu} 
                                                        & 33 & 0.40 & 0.37 \\
        & 2015 & $A_{LU}, A_{UL}, A_{LL}$ & \cite{CLAS:2015bqi}
                                                        & 389 & 1.24 & 1.47 \\
        & 2015 & $d\sigma_{UU}, \Delta\sigma_{LU}$ & \cite{Jo:2015ema} 
                                                        & 3828 & 0.93 & 0.98 \\
        & 2018 & 
        $
        \Delta\sigma_{LU}$ & \cite{CLAS:2018bgk} 
                                                        & 2465 & 1.96 & 1.93 \\
        & 2022 & $A_{LU}$ & \cite{CLAS:2022syx} 
                                                        & 1303 & 1.82 & 1.83 \\
        \midrule[0px]

        H1 
        & 2005 & $d\sigma_{UU}^{\gamma^*p}$ & \cite{Aktas:2005ty} 
                                                        & 8 & 1.59 & 0.44 \\
        & 2009 & $A_{C}, d\sigma_{UU}^{\gamma^*p}$ & \cite{H1:2009wnw} 
                                                        & 18 & 1.92 & 0.90 \\
        \midrule[0px]

        HallA 
        & 2015 & $d\sigma_{UU}, \Delta\sigma_{LU}$ & \cite{Defurne:2015kxq} 
                                                        & 600 & 0.64 & 0.63 \\
        & 2017 & $d\sigma_{UU}, \Delta\sigma_{LU}$ & \cite{Defurne:2017paw} 
                                                        & 806 & 1.15 & 1.60 \\
        & 2022 & $d\sigma_{UU}, \Delta\sigma_{LU}$ & \cite{JeffersonLabHallA:2022pnx} 
                                                        & 2640 & 1.23 & 1.40 \\
        \midrule[0px]

        HERMES 
        & 2001 & $A_{LU}$ & \cite{HERMES:2001bob} 
                                                        & 10 & 1.05 & 0.93 \\
        & 2006 & $A_{C}^{\cos\phi}$ & \cite{HERMES:2006pre} 
                                                        & 4 & 2.81 & 0.94 \\
        & 2008 & $A_{C}^{\cos(n\phi)}, A_{UT}^{\sin(\phi-\phi_s)\cos(n\phi)}$ & \cite{HERMES:2008abz} 
                                                        & 24 & 1.05 & 0.59 \\
        & 2009 & $A_{C}^{\cos(n\phi)}, A_{LU, \mathrm{I}}^{\sin(n\phi)}, A_{LU, \mathrm{DVCS}}^{\sin\phi}$ & \cite{HERMES:2009cqe} 
                                                        & 42 & 1.52 & 1.27 \\
        & 2010 & $A_{UL}^{\sin(n\phi)}, A_{LL}^{\cos(n\phi)}$ & \cite{HERMES:2010dsx} 
                                                        & 24 & 1.67 & 1.23 \\
        & 2011 & $A_{LT}$ & \cite{HERMES:2011bou} 
                                                        & 32 & 0.57 & 0.52 \\
        & 2012 & $A_{C}^{\cos(n\phi)}, A_{LU, \mathrm{I}}^{\sin(n\phi)}, A_{LU, \mathrm{DVCS}}^{\sin\phi}$ & \cite{HERMES:2012gbh} 
                                                        & 42 & 1.00 & 0.85 \\
        \midrule[0px]

        ZEUS 
        & 2008 & $d\sigma_{UU}^{\gamma^*p}$ & \cite{ZEUS:2008hcd} 
                                                        & 4 & 0.41 & 1.09 \\
        \midrule
        
        \textbf{Total} & & & & \textbf{12822} & \textbf{1.29} & \textbf{1.36} \\
        \bottomrule
        &&&&&\\
        \toprule
        \multicolumn{7}{c}{\textbf{NEUTRON TARGET}} \\
        \midrule
        CLAS & 2024 & $A_{LU}$ & \cite{CLAS:2024qhy} & 21 & 0.65 & 0.44 \\
        HallA & 2020 & $d\sigma_{UU}$ & \cite{Benali:2020vma} & 96 & 0.78 & 0.76\\
        \midrule                                              
        \textbf{Total} & & & & \textbf{117} & \textbf{0.76} & \textbf{0.70} \\
        \bottomrule
    \end{tabular}
    
\end{table}

\begin{figure}[!ht]
    \centering
    \includegraphics[width=1\textwidth]{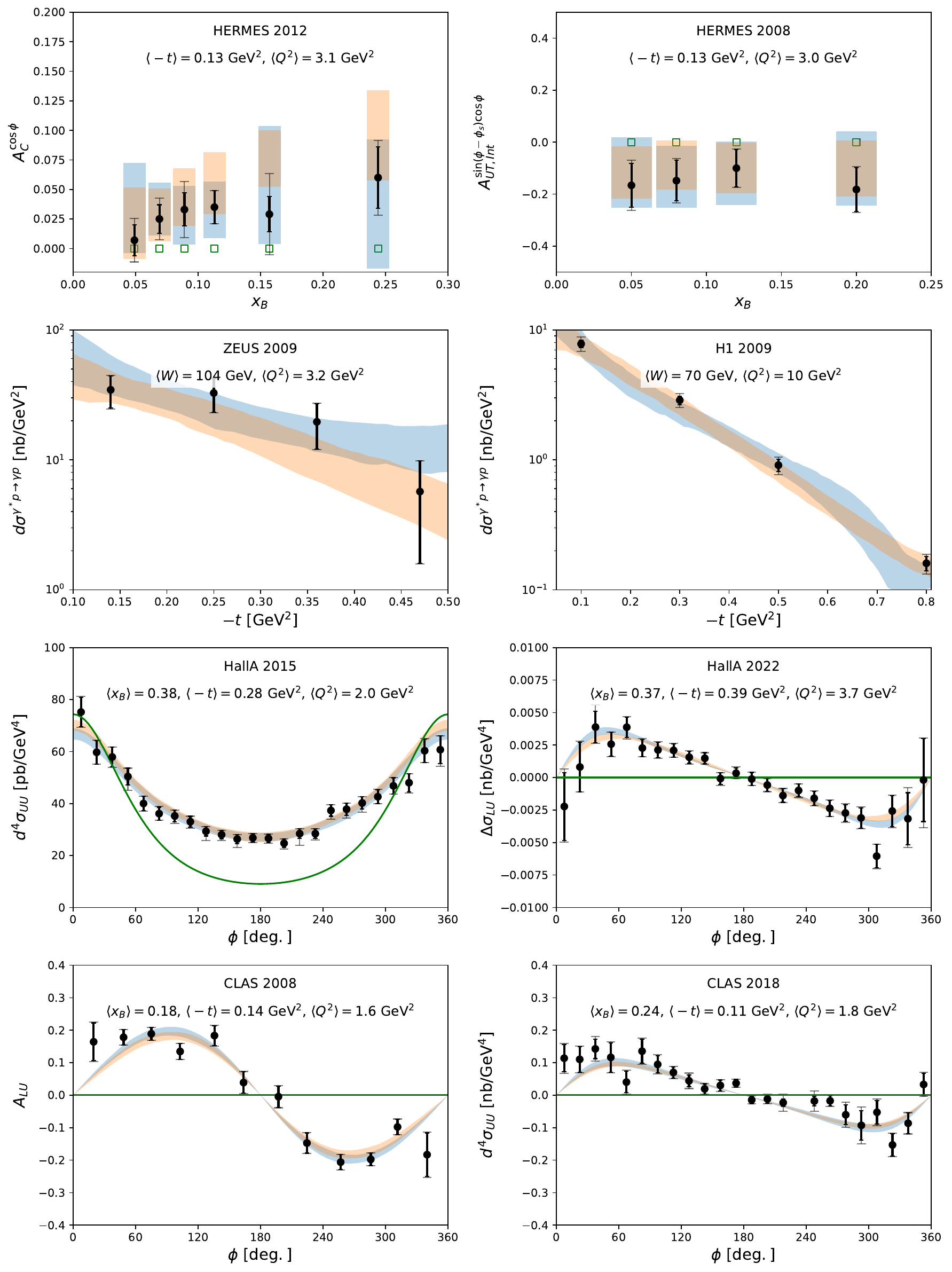}
    \caption{Example of the comparison of the fit results with the experimental data for proton target used in the fit. The pure BH contribution is denoted by the green points and curves. The resulting theory-agnostic and theory-augmented fits are shown by the blue and orange bands, respectively.}
    \label{fig:result:experiments}
\end{figure}

\begin{figure}[!ht]
    \centering
    \includegraphics[width=1\textwidth]{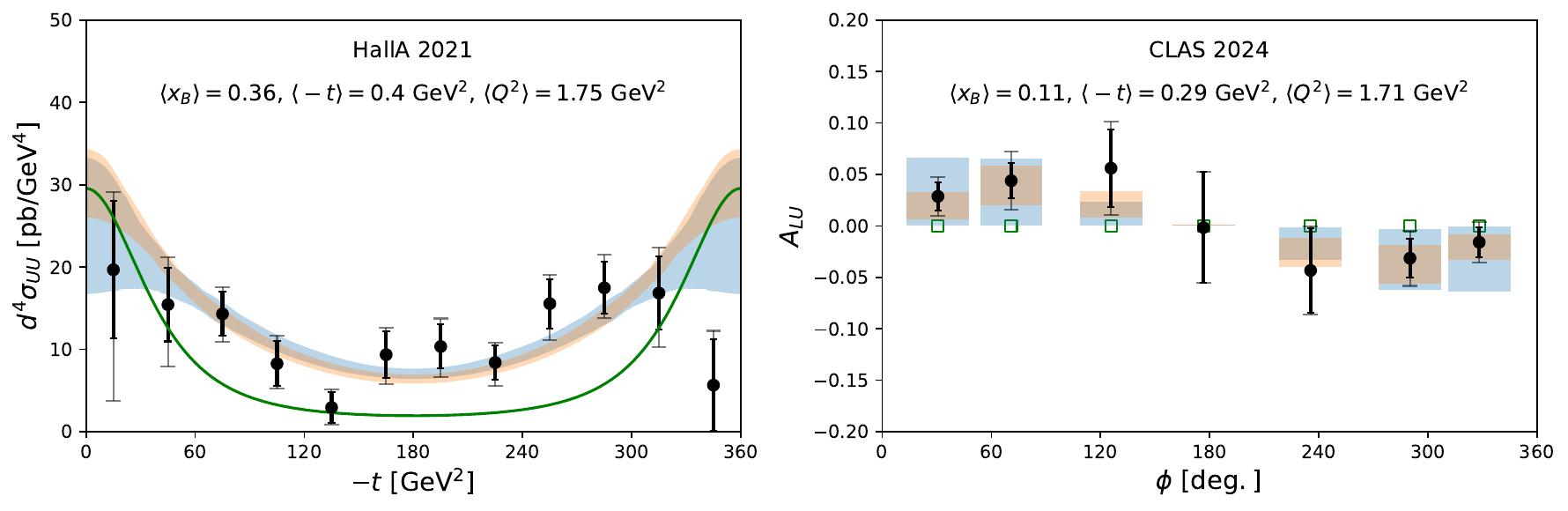}
    \caption{Example of the comparison of the fit results with the experimental data for neutron target used in the fit. For a further description, see the caption of Fig.~\ref{fig:result:experiments}.}
    \label{fig:result:experiments_neutron}
\end{figure}

A collection of plots comparing our fit results with the experimental data is presented in Figs.~\ref{fig:result:experiments} and~\ref{fig:result:experiments_neutron} for the proton and neutron targets, respectively. In cases such as the HERMES dataset, which in Fig.~\ref{fig:result:experiments} is presented as a function of $x_{B}$, the mean values of the other kinematic variables ($t$ and $Q^2$) vary from point to point. Therefore, a binned presentation is used instead of continuous lines and bands. In such cases, the mean values associated with the given datasets, displayed on top of each plot, are calculated as unweighted means over the data points and are provided only for guidance.

\begin{figure}[!ht]
    \centering
    \includegraphics[width=1\textwidth]{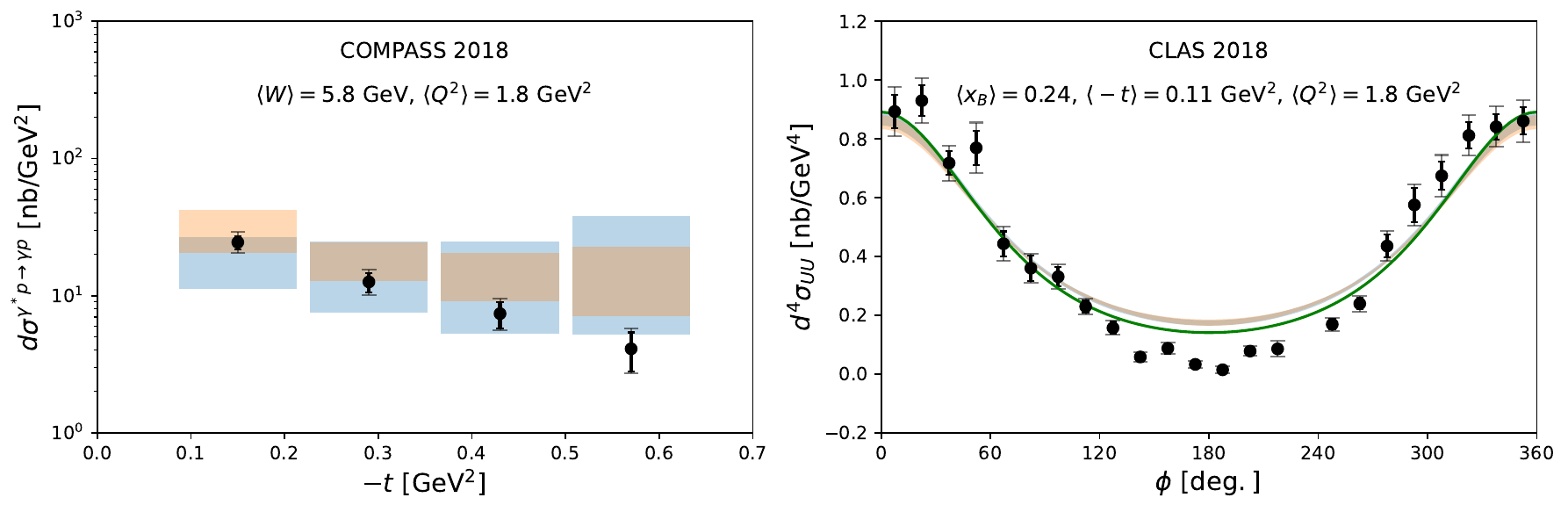}
    \caption{Example of the comparison of the fit results with the experimental data not used in the fit. For a further description, see the caption of Fig.~\ref{fig:result:experiments}.}
    \label{fig:result:experiments_not_used}
\end{figure}

We exclude two datasets from the fitting procedure. We plot them against the obtained results in Fig.~\ref{fig:result:experiments_not_used}. The COMPASS data~\cite{COMPASS:2018pup} are excluded because they are provided as cross sections integrated over broad kinematic bins, necessitating a corresponding treatment during the fit. At present, incorporating such multidimensional integration would significantly slow down the computation. Addressing this numerical challenge is deferred to future work, although Fig.~\ref{fig:result:experiments_not_used} demonstrates that these data offer valuable constraining power.

The other dataset removed from our analysis is the CLAS dataset published in 2018~\cite{CLAS:2018bgk}, but only for unpolarized cross sections (cross-section differences measured with various beam polarization states are kept in the fit). The reason for removing such a large portion of data is the bias it introduces into the fit. Currently, we are unable to describe these data, even in a local extraction. In Fig.~\ref{fig:result:experiments_not_used}, we show one of the problematic kinematic bins where the discrepancy is visually evident. Specifically, data points with very small uncertainties fall below the BH baseline for $\phi \sim \pi$.

\clearpage

\subsection{Compton form factors}

\begin{figure}[!ht]
    \centering
\includegraphics[width=1\textwidth]{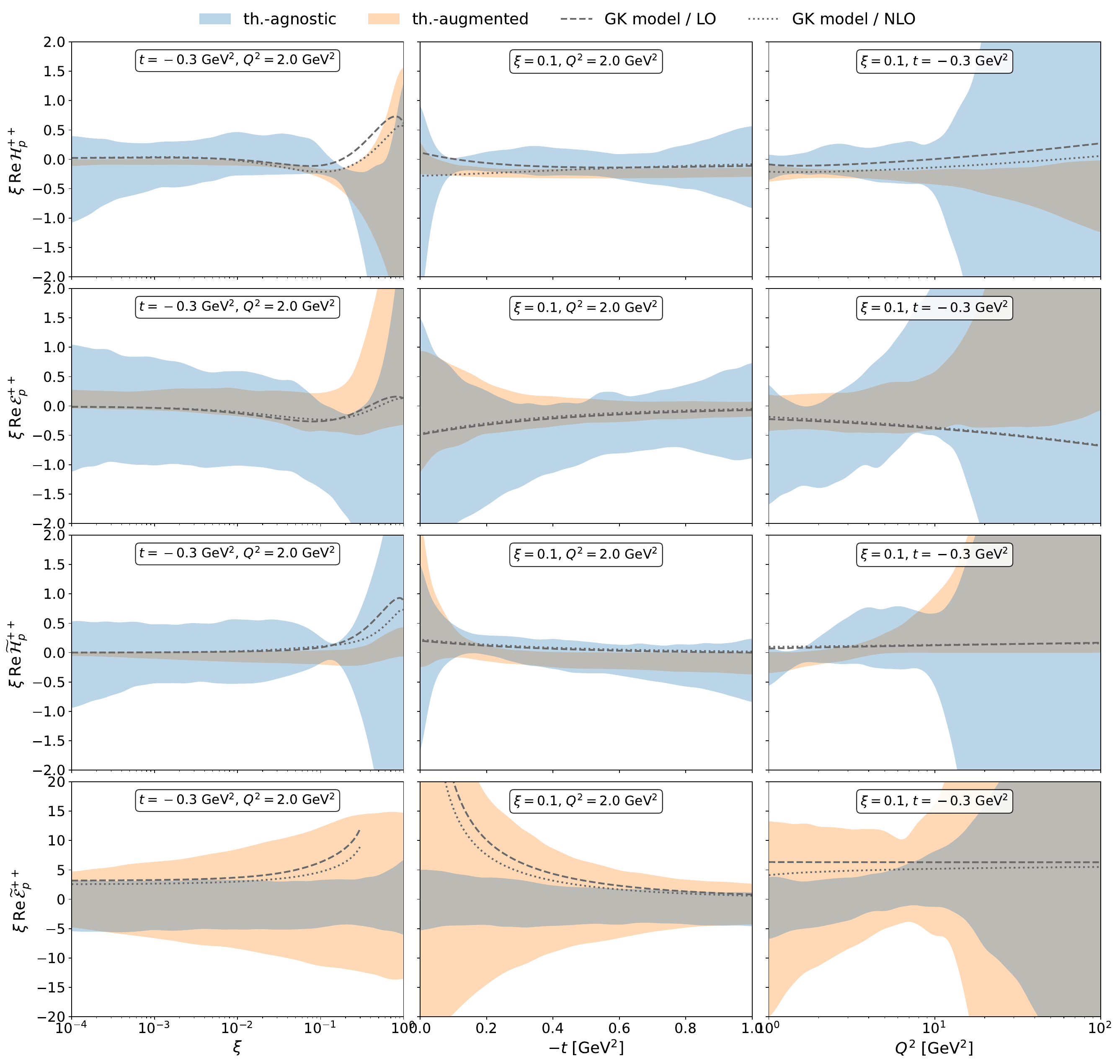}
    \caption{The real part of CFFs $\mathcal{H}^{++}_{p}$, $\mathcal{E}^{++}_{p}$, $\widetilde{\mathcal{H}}^{++}_{p}$, and $\widetilde{\mathcal{E}}^{++}_{p}$ extracted in this analysis from the proton data for the theory-agnostic (blue band) and theory-augmented (orange band) models. Predictions of the GK GPD model, evaluated with LO and NLO coefficient functions, are marked by dashed and dotted lines, respectively.}    \label{fig:result:cff_re_0}
\end{figure}

\begin{figure}[!ht]
    \centering
\includegraphics[width=1\textwidth]{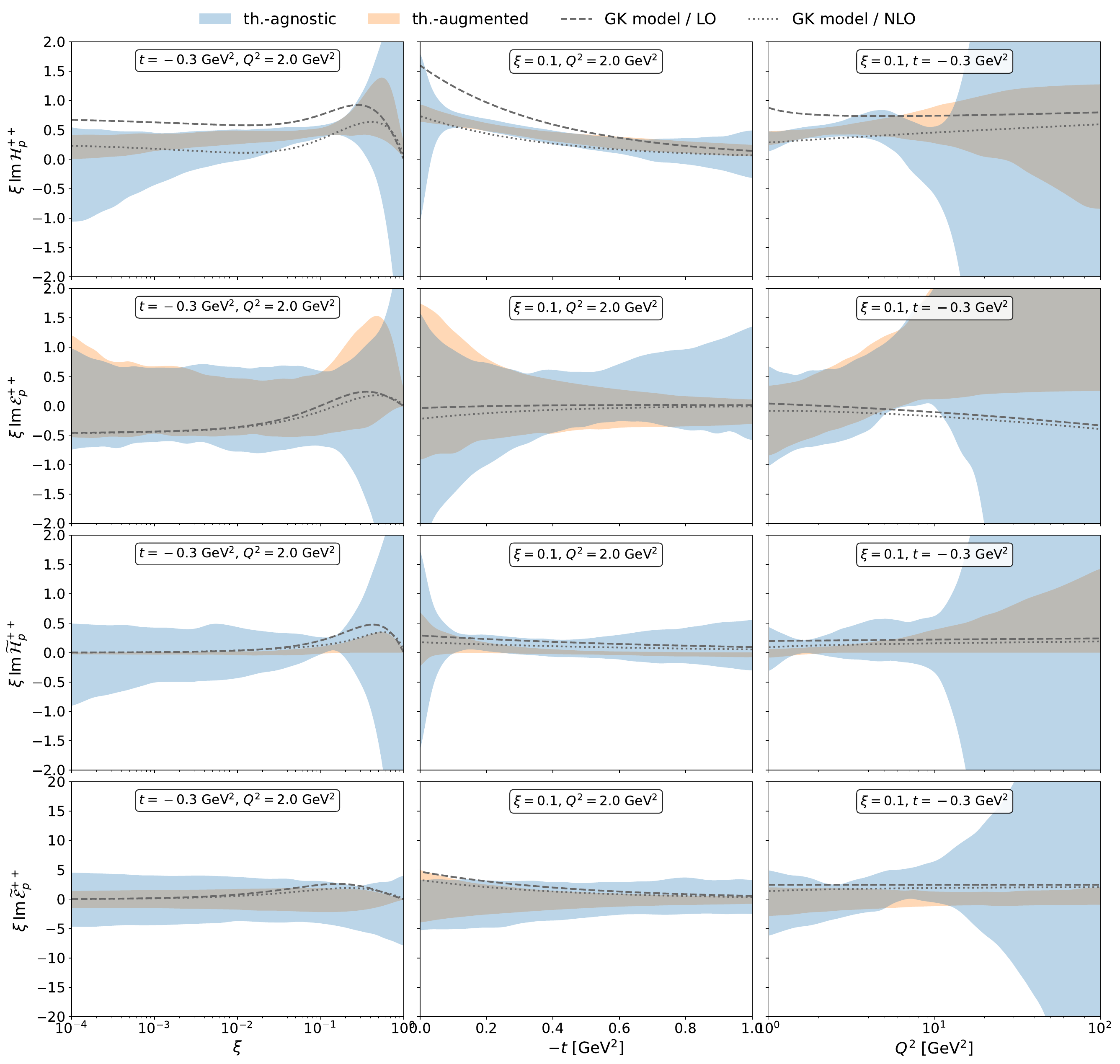}
    \caption{The imaginary part of CFFs $\mathcal{H}^{++}_{p}$, $\mathcal{E}^{++}_{p}$, $\widetilde{\mathcal{H}}^{++}_{p}$, and $\widetilde{\mathcal{E}}^{++}_{p}$ extracted in this analysis from the proton data. For a further description, see the caption of Fig.~\ref{fig:result:cff_re_0}.}
    \label{fig:result:cff_im_0}
\end{figure}

\begin{figure}[!ht]
    \centering
\includegraphics[width=1\textwidth]{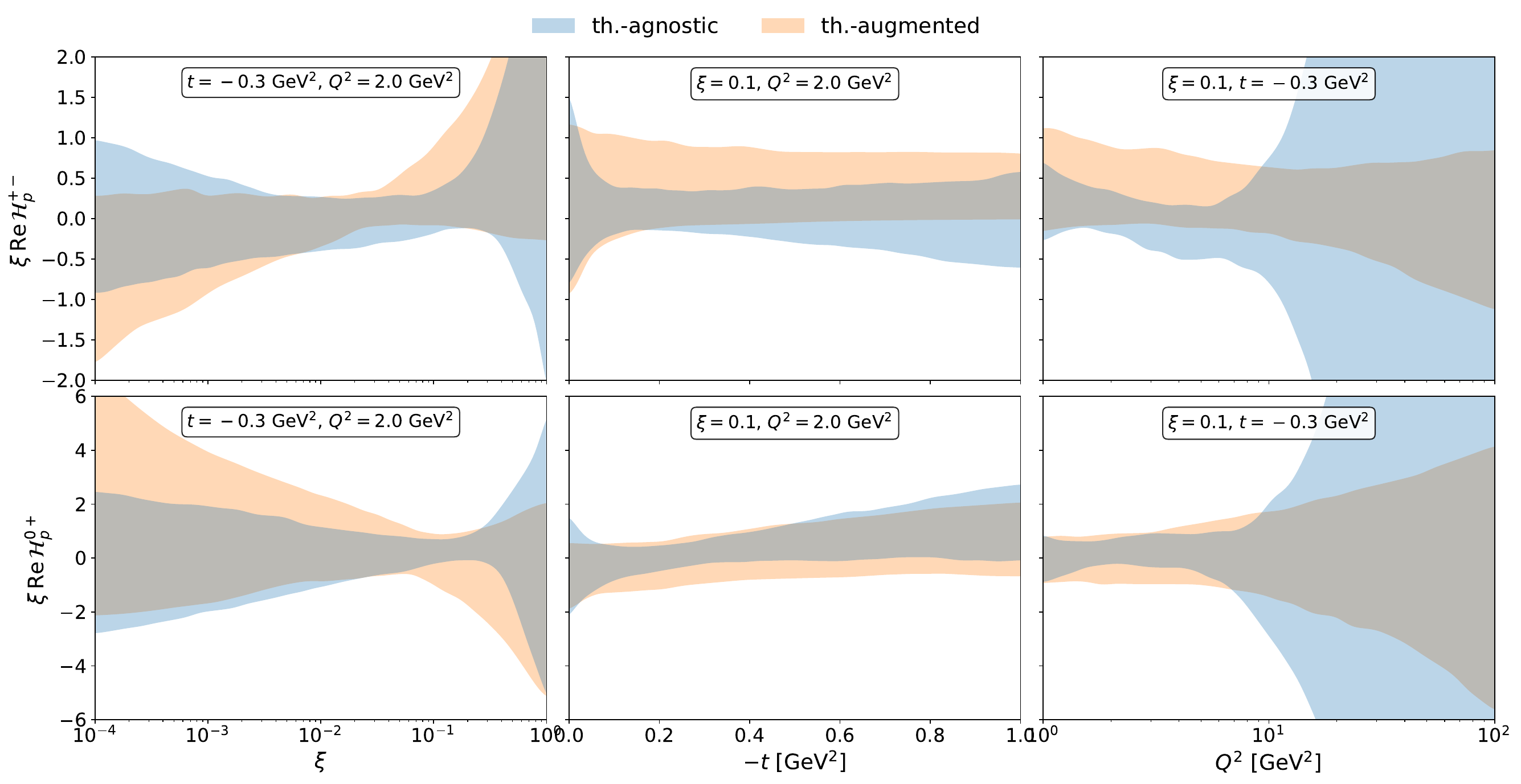}
    \caption{The real part of CFFs $\mathcal{H}^{+-}_{p}$ and $\mathcal{H}^{0+}_{p}$ extracted in this analysis from the proton data. For a further description, see the caption of Fig.~\ref{fig:result:cff_re_0}.}
    \label{fig:result:cff_re_1}
\end{figure}

\begin{figure}[!ht]
    \centering
\includegraphics[width=1\textwidth]{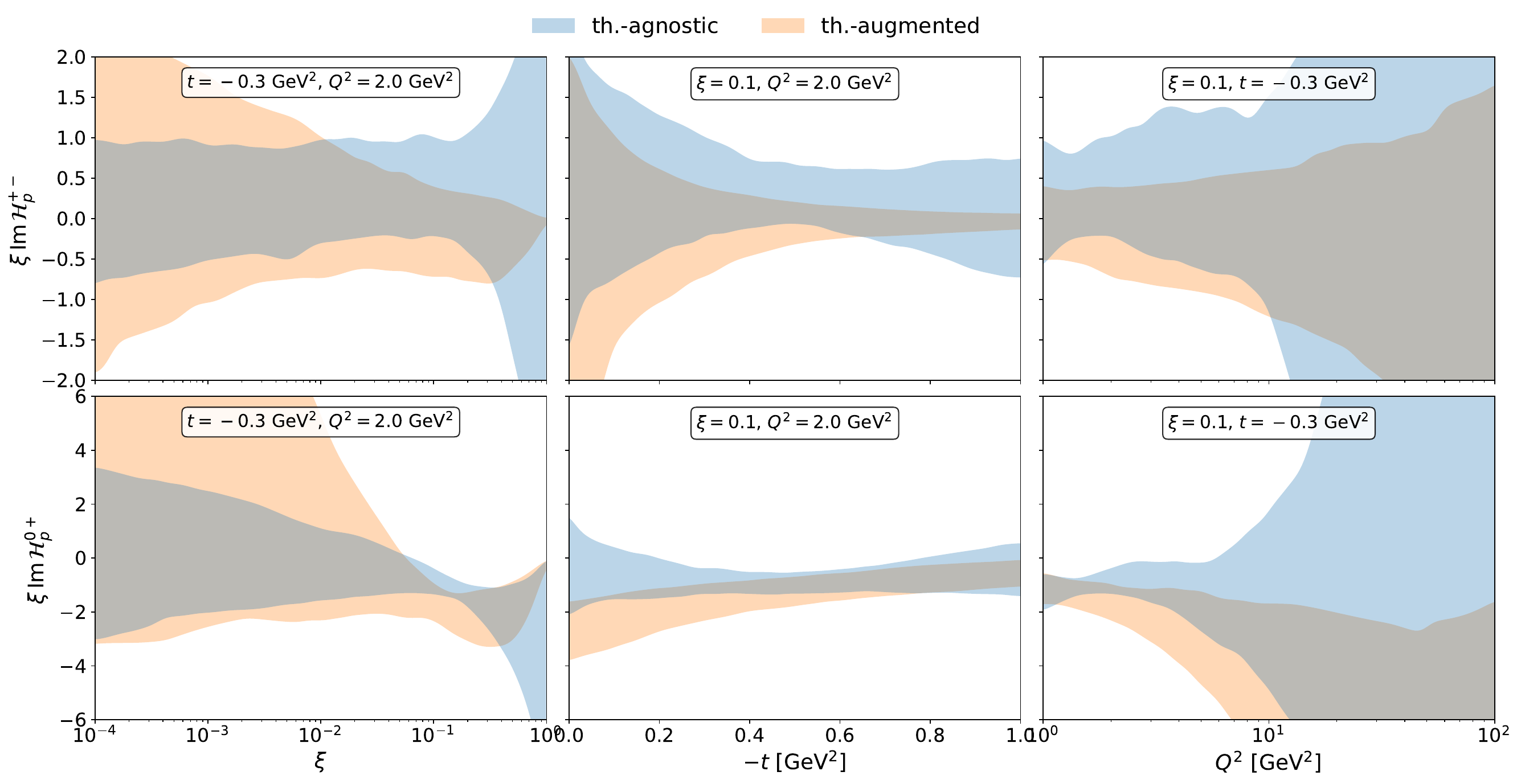}
    \caption{The imaginary part of CFFs $\mathcal{H}^{+-}_{p}$ and $\mathcal{H}^{0+}_{p}$ extracted in this analysis from the proton data. For a further description, see the caption of Fig.~\ref{fig:result:cff_re_0}.}
    \label{fig:result:cff_im_1}
\end{figure}

\begin{figure}[!ht]
    \centering
\includegraphics[width=1\textwidth]{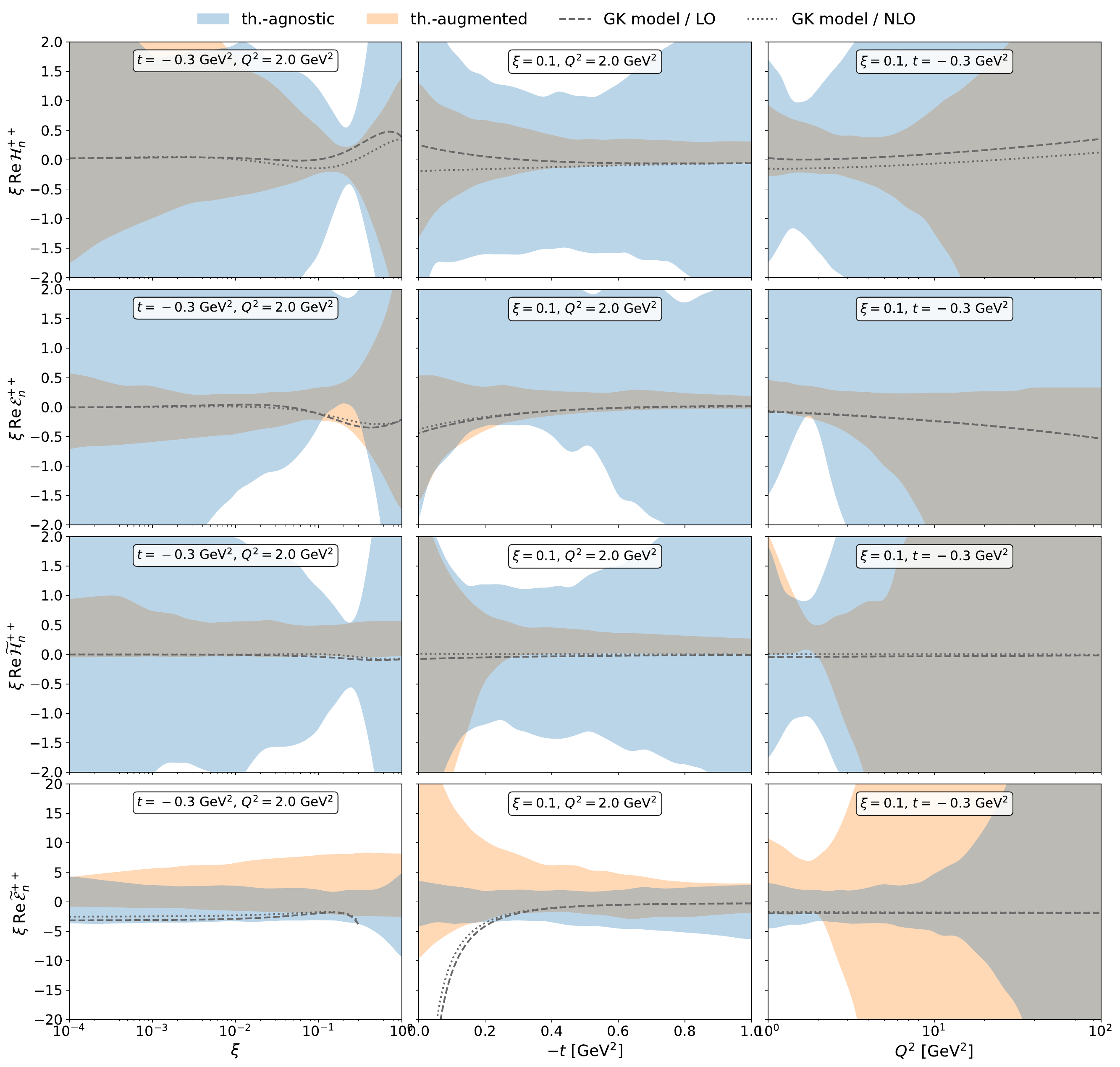}
    \caption{The real part of CFFs $\mathcal{H}^{++}_{n}$, $\mathcal{E}^{++}_{n}$, $\widetilde{\mathcal{H}}^{++}_{n}$, and $\widetilde{\mathcal{E}}^{++}_{n}$ extracted in this analysis from the neutron data. For a further description, see the caption of Fig.~\ref{fig:result:cff_re_0}.}    \label{fig:result:cff_re_0_neutron}
\end{figure}

\begin{figure}[!ht]
    \centering
\includegraphics[width=1\textwidth]{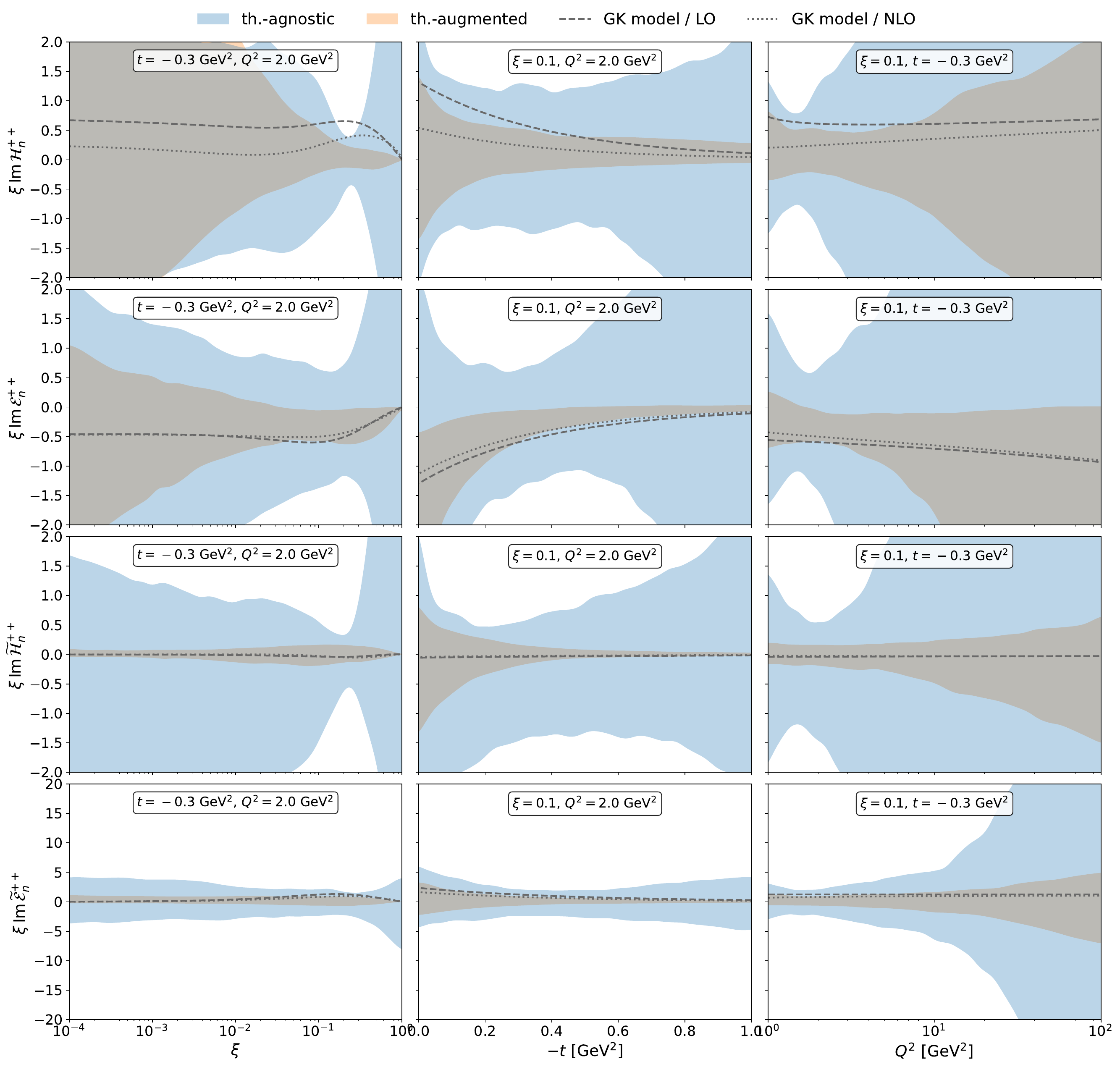}
    \caption{The imaginary part of CFFs $\mathcal{H}^{++}_{n}$, $\mathcal{E}^{++}_{n}$, $\widetilde{\mathcal{H}}^{++}_{n}$, and $\widetilde{\mathcal{E}}^{++}_{n}$ extracted in this analysis from the neutron data. For a further description, see the caption of Fig.~\ref{fig:result:cff_re_0}.}
    \label{fig:result:cff_im_0_neutron}
\end{figure}

For the proton, the distributions of the extracted CFFs $\mathcal{H}^{++}_{p}$, $\mathcal{E}^{++}_{p}$, $\widetilde{\mathcal{H}}^{++}_{p}$, and $\widetilde{\mathcal{E}}^{++}_{p}$ are shown in Figs.~\ref{fig:result:cff_re_0} and~\ref{fig:result:cff_im_0} for their real and imaginary parts, respectively\footnote{Hereafter, the subscripts $p$ and $n$ denote CFFs extracted for the proton and neutron targets, respectively.}. The corresponding distributions for the helicity-flip CFFs $\mathcal{H}^{0+}_{p}$ and $\mathcal{H}^{+-}_{p}$ are presented in Figs.~\ref{fig:result:cff_re_1} and~\ref{fig:result:cff_im_1}. For the neutron, the extracted CFFs $\mathcal{H}^{++}_{n}$, $\mathcal{E}^{++}_{n}$, $\widetilde{\mathcal{H}}^{++}_{n}$, and $\widetilde{\mathcal{E}}^{++}_{n}$ are shown in Figs.~\ref{fig:result:cff_re_0_neutron} and~\ref{fig:result:cff_im_0_neutron}. The differences $\mathcal{H}^{++}_{p} - \mathcal{H}^{++}_{n}$ and $\mathcal{E}^{++}_{p} - \mathcal{E}^{++}_{n}$ are presented and discussed in detail in Sect.~\ref{sec:result:p_minus_n}.

Overall, for both proton and neutron targets, we observe good agreement between the results obtained from the two models. As expected, the uncertainties yielded by the theory-augmented model are typically smaller, particularly in regions poorly constrained by experimental data. The only exception is the real part of the poorly constrained CFF $\widetilde{\mathcal{E}}^{++}$, which exhibits larger uncertainties in the theory-augmented approach. One possible explanation is that evaluating the real part requires integrating the corresponding imaginary part over the full range of $\xi$. This integration inherently propagates uncertainties from unmeasured kinematic regions into the error band of the real part, an effect that appears to be particularly severe in this case.

In the plots, we also include a comparison with the GK GPD model, which was used to evaluate the CFFs utilizing either LO or NLO coefficient functions. This model only includes the GPDs $H$, $E$, $\widetilde{H}$, and $\widetilde{E}$, and lacks a model for the D-term. We also use this model with its built-in forward evolution in $\mu^2$. Looking specifically at the dominant CFF $\mathcal{H}^{++}$, we observe that the model exhibits better agreement with our extraction when evaluated using NLO coefficient functions. This highlights the importance of higher-order corrections in the extraction of GPDs.

The helicity-flip CFFs $\mathcal{H}^{0+}$ and $\mathcal{H}^{+-}$ are generally consistent with zero, with the possible exception of a negative signal for $\mathcal{H}^{0+}$ at high $\xi$. The emergence of such a negative value is particularly interesting given the strong sensitivity of this amplitude to higher-twist effects. From the evaluation performed in Ref.~\cite{Braun:2025xlp} and based on the GK model, the sign of the real and imaginary parts of $\mathcal{H}^{0+}$ seems to depend heavily on the kinematics and the power truncation. A non-vanishing $\mathcal{H}^{0+}$ may indicate the increasing relevance of higher-twist effects, suggesting a breakdown of the strict leading-twist approximation and emphasizing the need to systematically include such corrections in future phenomenological studies of GPDs. When discussing higher-twist corrections, it is important to emphasize that CFFs are not constrained to vanish in the limit $t \to 0$~\cite{Braun:2014sta}.

\begin{figure}[!ht]
    \centering
\includegraphics[width=1\textwidth]{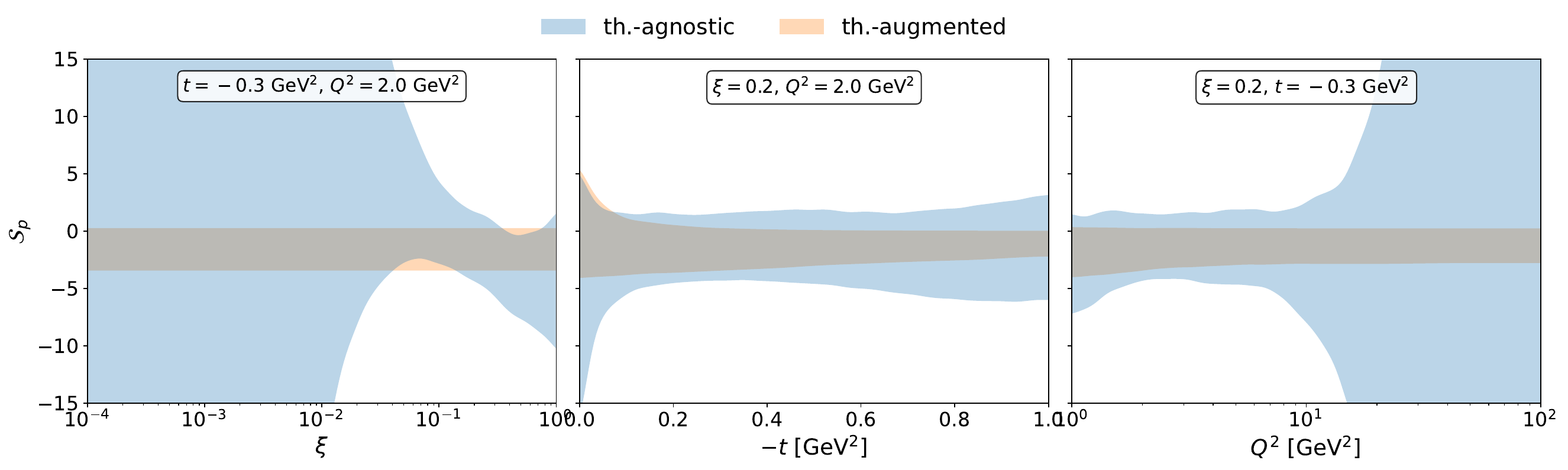}
    \caption{The subtraction constant extracted in this analysis from the proton data. For a further description, see the caption of Fig.~\ref{fig:result:cff_re_0}.}
    \label{fig:result:cff_sc}
\end{figure}

\begin{figure}[!ht]
    \centering
\includegraphics[width=1\textwidth]{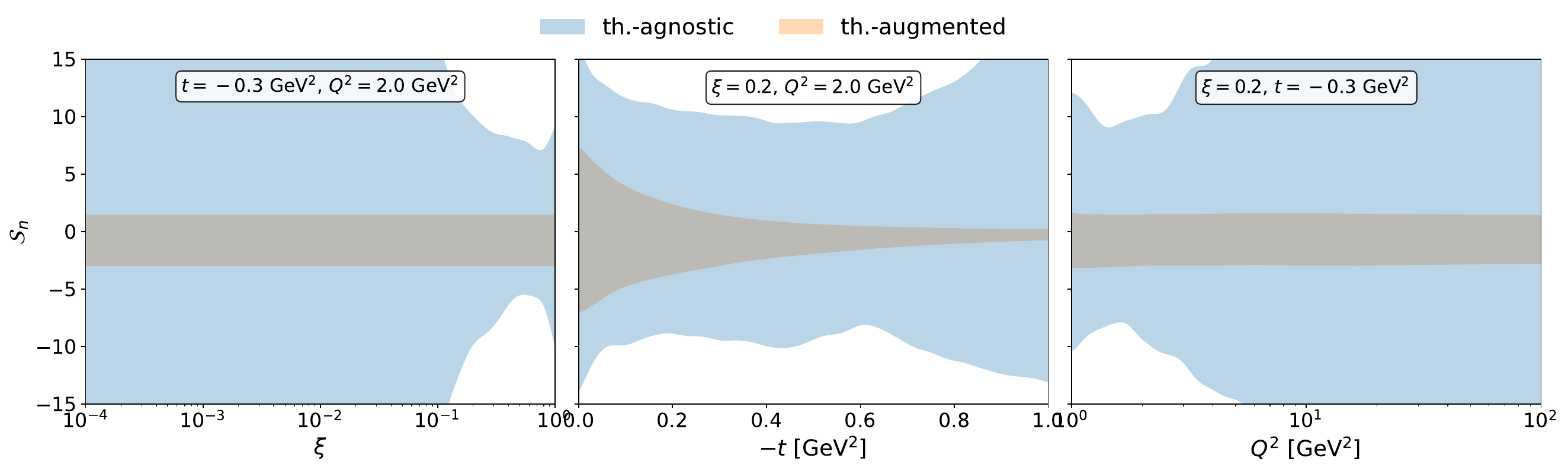}
    \caption{The subtraction constant extracted in this analysis from the neutron data. For a further description, see the caption of Fig.~\ref{fig:result:cff_re_0}.}
    \label{fig:result:cff_sc_neutron}
\end{figure}

The distribution of the subtraction constant extracted from the proton data is shown in Fig.~\ref{fig:result:cff_sc}. For the theory-agnostic model, this quantity is evaluated using the dispersion relation after the fit, with the real and imaginary parts of the CFF $\mathcal{H}^{++}$ constrained independently by the DVCS data. In contrast, in the theory-augmented model the dispersion relation is imposed directly in the fit, such that the subtraction constant is constrained simultaneously with the imaginary part of $\mathcal{H}^{++}$. Analogous results obtained from the analysis of neutron data are shown in Fig.~\ref{fig:result:cff_sc_neutron}.

Consistent with the CFF results, we observe that the uncertainties associated with the theory-agnostic model are larger. For this model, we evaluate the dispersion relation as a function of $\xi$ to serve as a consistency check for its validity in the present extraction. The asymptotic value of the subtraction constant at $t = 0$ and $Q^2 \to \infty$ is of particular importance because of its direct relation to the energy-momentum form factor $C$ (see, e.g., Ref.~\cite{Dutrieux:2021nlz}). In our analysis, we extract $\Delta_0 = -0.69^{+6.92}_{-3.13}$ with the theory-augmented model. This value is consistent with our previous extraction~\cite{Dutrieux:2021nlz}, which was similarly based on a global data analysis, but restricted to measurements published prior to 2016. Qualitatively, however, the inclusion of recent data does not significantly alter the extraction of the subtraction constant, primarily because these data lack strong sensitivity to the real part.


For the neutron data, the uncertainties in the extracted values of the asymptotic term within the subtraction constant are predictably larger than those for the proton. However, this increase is surprisingly mild, and the uncertainties do not scale as poorly as one would naively expect based solely on the limited number of available data points. Specifically, we obtain $-0.86^{+7.25}_{-5.59}$ with the theory-augmented model. This disproportionate sensitivity of the neutron data to the subtraction constant may stem from three primary physical advantages. First, the suppression of the BH background in neutron scattering significantly enhances the relative strength of the pure GPD signal compared to proton extractions. Second, assuming flavour symmetry for the D-term and therefore the subtraction constant (i.e., $\mathcal{S}_p \approx \mathcal{S}_n$), its relative contribution to the neutron amplitude is amplified, because the imaginary part of the CFF is, on average, larger for the proton ($\mathrm{Im}\mathcal{H}^{++}_p > \mathrm{Im}\mathcal{H}^{++}_n$). Finally, neutron observables are naturally more sensitive to the GPD $E$, which independently encodes D-term contributions. All these factors motivate further measurements on the neutron.


\clearpage

\subsection{Subtraction constant vs. lattice-QCD and CSM calculations}

\begin{figure}[!ht]
    \centering
\includegraphics[width=\textwidth]{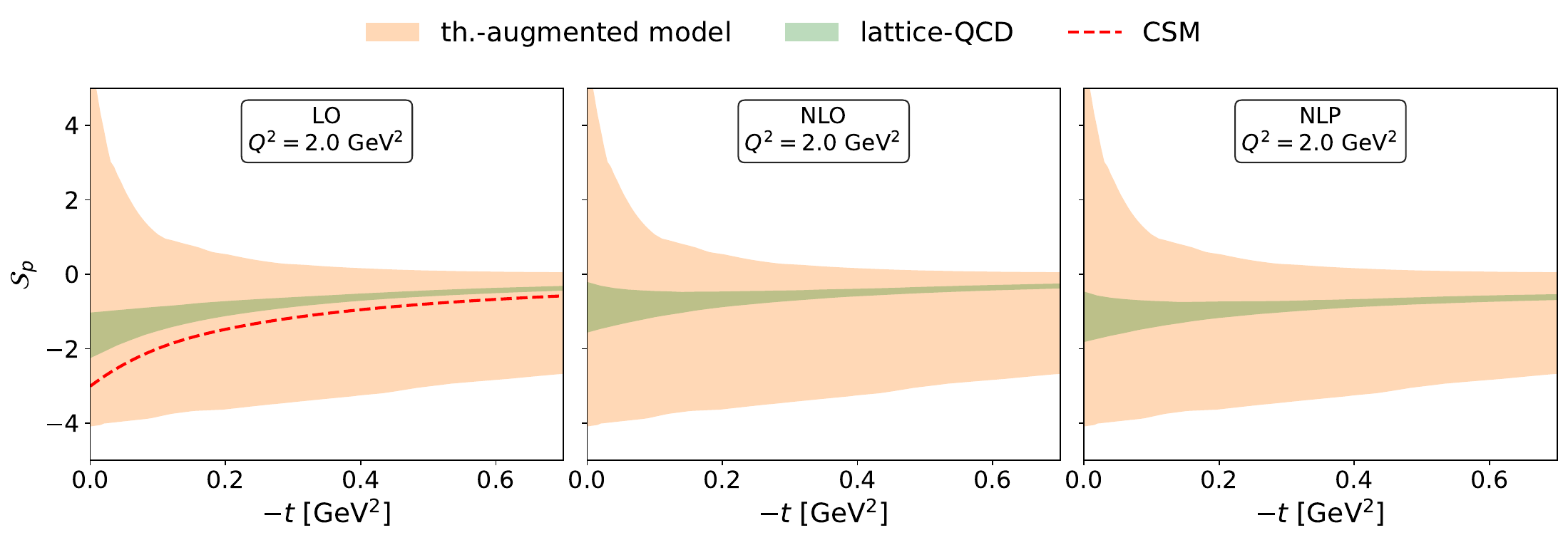}
    \caption{The subtraction constant extracted in this analysis compared to prediction from lattice-QCD EMT form factors coming from Ref.~\cite{Hackett:2023rif}. Note that lattice artefacts and truncation effects are not included in the uncertainty band.}
    \label{fig:result:sc_vs_latt}
\end{figure}

From the previous section, we are now in a position to compare our extraction of the subtraction constant defined in Eq.~\eqref{eq:DVCSDispersionRelation} with results obtained from \emph{ab initio} methods, in particular lattice QCD. 
There are no lattice-QCD data today related to the subtraction constant \emph{per se}. 
However, the subtraction constant can be computed using lattice-QCD extractions of the generalized form factors. 
Indeed, combining the approximate expressions for the quark and gluon subtraction constants at NLO and NLP (but without corrections of the type $t/Q^2 \alpha_s$ or $M^2/Q^2 \alpha_s$) given respectively in Refs.~\cite{Dutrieux:2024bgc} and~\cite{Martinez-Fernandez:2025rcg,Martinez-Fernandez:2025jvk}, we get:
\begin{align}
    \label{eq:SubtractionConstantApprox}
    \mathcal{S} & = \sum_q e_q^2 \mathcal{S}^q + \mathcal{S}^g\,, \\
    \mathcal{S}^q & \simeq \left(4-\frac{4}{9}\frac{\alpha_s C_F}{4\pi} - \frac{4}{3}\frac{t}{\mathbb{Q}^2}\right)d_1^q(t) - 4\frac{M^2}{\mathbb{Q}^2}c_0\left(A^q_{1,0}(t) + \frac{t}{4M^2}B^q_{1,0}(t)\right)\,, \\
     \mathcal{S}^g & \simeq \frac{\sum_q e_q^2 \alpha_s T_F}{4\pi}\left(-\frac{172}{9} d_1^g(t) \right)\,,
\end{align}
where $\mathbb{Q}^2 = Q^2 + t$, $C_F = 4/3 $, $T_F = 1/2 $ and $c_0\simeq 0.860$.
We have considered only the first moments $d_1(t)$ in the expansion of the underlying Polyakov-Weiss $D$-term \cite{Polyakov:1999gs} on a Gegenbauer $C^{3/2}$ basis for quarks and  $C^{5/2}$ for gluons \cite{Dutrieux:2021nlz}: 
\begin{align}
    \label{eq:DTermDef}
    D^q(z,t) = (1-z^2)\sum_{n=0}^\infty d^q_{2n+1}(t) C_{2n+1}^{3/2}(z), \\
    D^g(z,t) = \frac{3}{2}(1-z^2)^2 \sum_{n=0}^\infty d^g_{2n+1}(t) C_{2n}^{5/2}(z).
\end{align}
$A^q_{1,0}(t)$ and $B^q_{1,0}(t)$ are two of the Energy-Momentum Form Factors defined as\footnote{Note that the Polyakov-Weiss $D$-term $D(z,t)$ has nothing to do with the EMT Form Factor $D(t)$. The notation is unfortunate but standard in the literature.}:
\begin{align}
    \label{eq:EMTDef}
    \bra{p',s'}T^q_{\mu\nu}\ket{p,s} = \bar{u}(p',s') & \left[ \frac{P_\mu P_\nu}{M} A_{1,0}^q(t) + \frac{\Delta_\mu \Delta_\nu -\eta_{\mu\nu} \Delta^2}{M} C^q(t) + M\eta_{\mu\nu} \bar{C}^q(t) \right. \nonumber \\
    & \left. + \frac{P_{\{\mu}i\sigma_{\nu\} \rho}\Delta^\rho}{4M}\left(A_{1,0}^q(t) + B_{1,0}^q(t) \right) + + \frac{P_{[\mu}i\sigma_{\nu] \rho}\Delta^\rho}{4M}D^q(t)  \right],
\end{align}
where $\{\dots\}$ indicates that the indices are symmetrised while $[ \dots ]$ that they are antisymmetrised. 
Knowing that $d_1$ is connected to the EMT form factor $C$ through
\begin{align}
    \label{eq:d1vsC}
    d_1^i(t) = 5C^i(t),
\end{align}
one can exploit both the lattice QCD results for the EMT form factors obtained in Ref.~\cite{Hackett:2023rif} and the Continuum Schwinger Method (CSM) one obtained in Ref.~\cite{Yao:2024ixu}. 
The comparison is presented in Fig.~\ref{fig:result:sc_vs_latt}.
Some caveats should nevertheless be mentioned. 
On the lattice side, only a single ensemble was considered, preventing the extrapolation of the results to the continuum limit. 
The pion mass is not too high, but is still $20\%$ higher than the physical one. 
These limitations should be understood as additional uncertainties \emph{not} taken into account in the uncertainty band presented here.
However, we do not believe these additional sources of uncertainties will inflate the uncertainty band so that it becomes comparable to the phenomenological one.
On the CSM side, the computations are done in a specific scheme, preventing us to compute the NLO corrections.
We thus present the CSM results only in the LO-LP case.
We also do not have the associated uncertainties at our disposal, but again, it is unlikely that the associated uncertainties are comparable with the experimental one. 
Note that the total (\emph{i.e.} summed over quarks and gluons contributions) EMT form factors between the two methods are in agreement, as it should since this quantity is scheme and scale invariant. 

In order to propagate the lattice uncertainties, we generated a correlated set of 500 replicas from the dipole fit parameters obtained in Ref.~\cite{Hackett:2023rif}.
Compared to the lattice-QCD error band, one can emphasise the sizable effect of the $C^g$ form factors, ensuring that the predictions at NLO are significantly closer to zero, as gluons contribute with a relative minus sign with respect to quarks.
This effect is slightly compensated by NLP corrections, which emphasise a bit the quarks sector. 
However, due to the lack of knowledge of the real part of the CFF, the uncertainty band produced by our extraction is much larger than the lattice one in all scenarios, and much larger than the difference between CSM and lattice computations.
A better determination of the real part of the CFF might come from the positron program at Jefferson Laboratory~\cite{Dutrieux:2021ehx}. 
Nevertheless, to the best of our knowledge, this is the first quantitative comparison between a DVCS observable and a prediction based on lattice QCD including both NLO and NLP corrections.

\subsection{Universality check through timelike Compton scattering}

\newcommand{\conj}{\operatorname{conj}}

With the CFFs extracted for DVCS, in the spirit of the analysis outlined in Ref.~\cite{Grocholski:2019pqj}, we can now evaluate observables for the TCS channel and compare them with the measurements by CLAS~\cite{CLAS:2021lky}. To do this, we use the known relations between the DVCS and TCS amplitudes~\cite{Mueller:2012sma}, which differ depending on the order in $\alpha_s$ considered:
\begin{eqnarray}
{^T{\cal H^{++}}}  &\stackrel{\rm LO}{=}&  \conj\left(^S{\cal H^{++}}\right) \,,
\\
{^T\widetilde{\cal H}^{++}}  &\stackrel{\rm LO}{=}& -\conj\left(^S\widetilde{\cal H}^{++}\right) \,,
\\
{^T{\cal H^{++}}}  &\stackrel{\rm NLO}{=}&  \conj\left(^S{\cal H^{++}}\right) - i\pi\, {Q}^2\frac{\partial}{\partial {Q}^2} \conj\left(^S{\cal H^{++}}\right) \,,
\\
{^T\widetilde{\cal H}^{++}}  &\stackrel{\rm NLO}{=}& -\conj\left(^S\widetilde{\cal H}^{++}\right) + i\pi\, {Q}^2\frac{\partial}{\partial {Q}^2} \conj\left(^S\widetilde{\cal H}^{++}\right)\,,
\end{eqnarray}
where $\conj(\cdot)$ denotes complex conjugation, and the superscripts $T$ and $S$ on the CFFs indicate the timelike (TCS) and spacelike (DVCS) cases, respectively. We use the corresponding relations for the CFFs ${\cal E}^{++}$ and $\widetilde{\cal E}^{++}$. At present, we refrain from studying the impact of helicity-flip amplitudes, leaving this development for future studies.

\begin{figure}[!ht]
    \centering
    \includegraphics[width=1\textwidth]{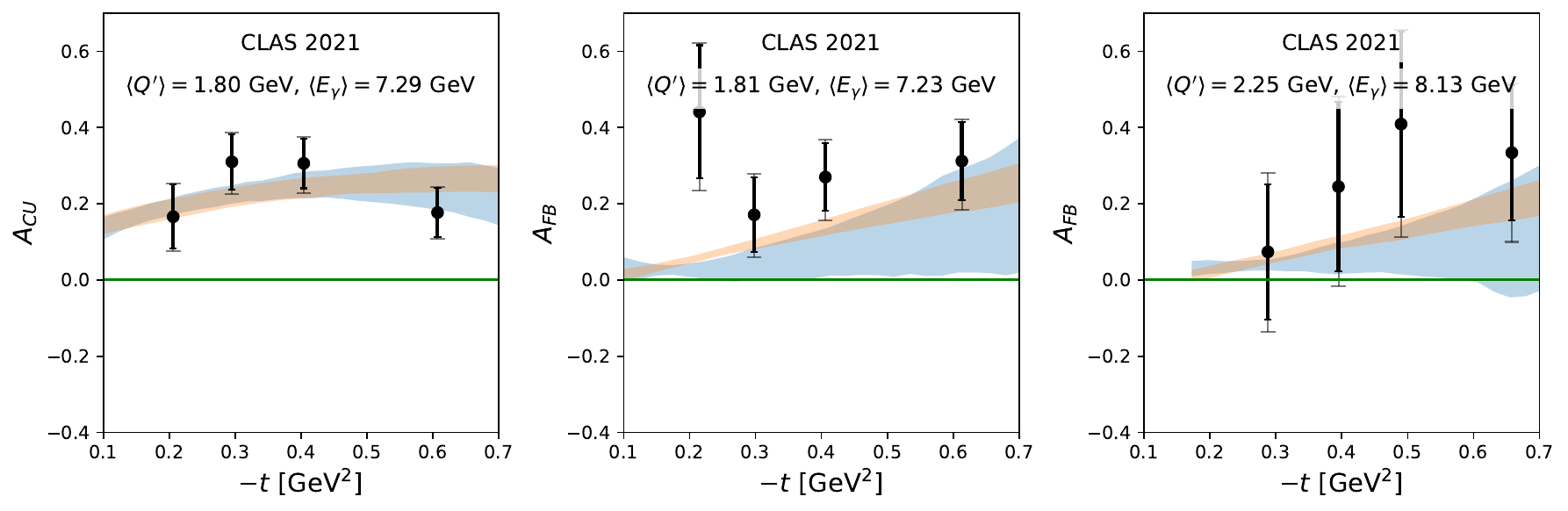}
    \includegraphics[width=1\textwidth]{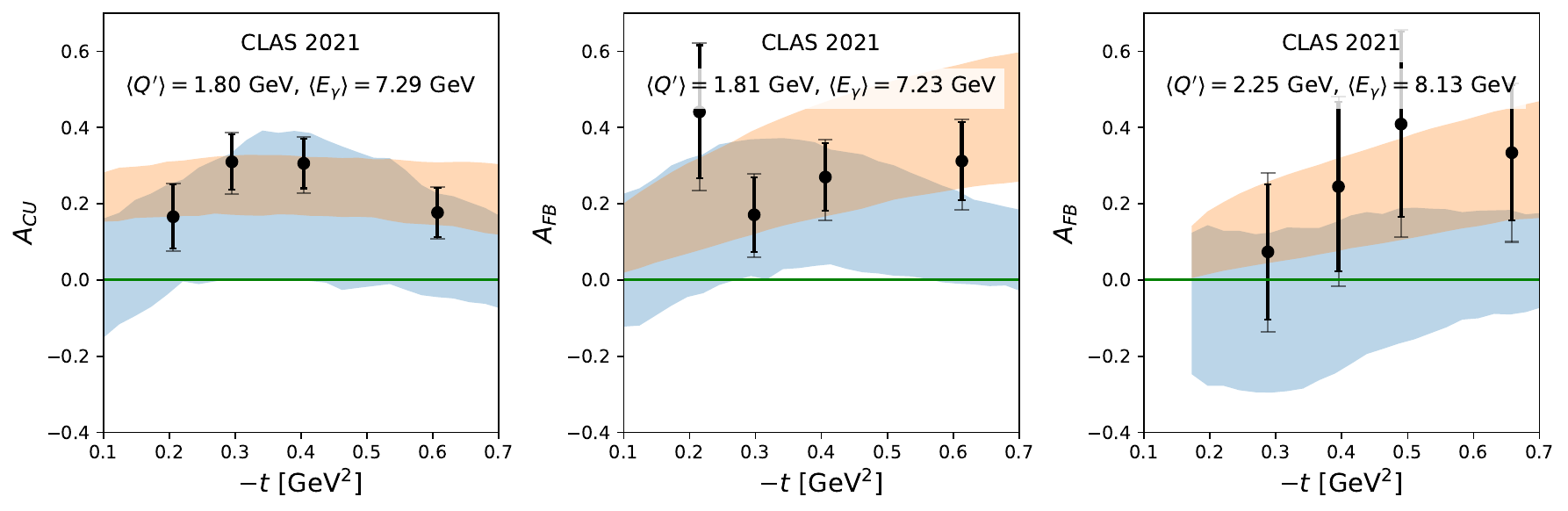}
    \caption{TCS observables evaluated from the extracted DVCS amplitudes using LO (first row) and NLO (second row) relations between the amplitudes of both processes, including a comparison with CLAS data~\cite{CLAS:2021lky}.}
    \label{fig:result:experiments_tcs}
\end{figure}

The results are presented in Fig.~\ref{fig:result:experiments_tcs} for both the LO and NLO relations. The uncertainties in the NLO case are larger, as this evaluation involves a derivative with respect to $Q^2$, along with a mixing of real and imaginary parts known with varying levels of precision. In general, we observe good agreement between the extracted DVCS amplitudes and the currently available, yet still exploratory, TCS data. Interestingly, the data seem to prefer the NLO predictions. This preference is particularly visible for the theory-augmented model, where a comparison between the experimental data and the median of the predictions yields $\chi^2 / N_{\mathrm{points}} = 0.91$ at LO and $\chi^2 / N_{\mathrm{points}} = 0.47$ at NLO. The plots also demonstrate a substantial potential for TCS data to constrain GPD information in precise extractions of GPDs, in particular beyond LO.

\subsection{Isovector CFFs}
\label{sec:result:p_minus_n}

Having extracted the CFFs independently for both the proton and the neutron, we are now in a position to evaluate their differences, such as $\mathcal{H}^{++}_{p} - \mathcal{H}^{++}_{n}$ and $\mathcal{E}^{++}_{p} - \mathcal{E}^{++}_{n}$. Under the assumption of exact isospin symmetry, comparing proton and neutron observables provides a direct pathway to flavor separation. Because the electromagnetic charges of the up and down quarks weight the proton and neutron GPDs differently, this difference effectively isolates the non-singlet (isovector) quark combinations, specifically the $u - d$ contribution. Furthermore, because gluons couple identically to both nucleons, all flavor-singlet and gluonic contributions cancel out exactly in this subtraction. Consequently, the proton-neutron CFF difference serves as an exceptionally clean probe of valence quark dynamics, free from gluon contamination. It is also an ideal observable for comparison with lattice-QCD results (see for instance Ref.~\cite{Dutrieux:2026grg}), since the evaluation of this difference can be achieved with significantly better control over systematic uncertainties.

\begin{figure}[!ht]
    \centering
    \includegraphics[width=1\textwidth]{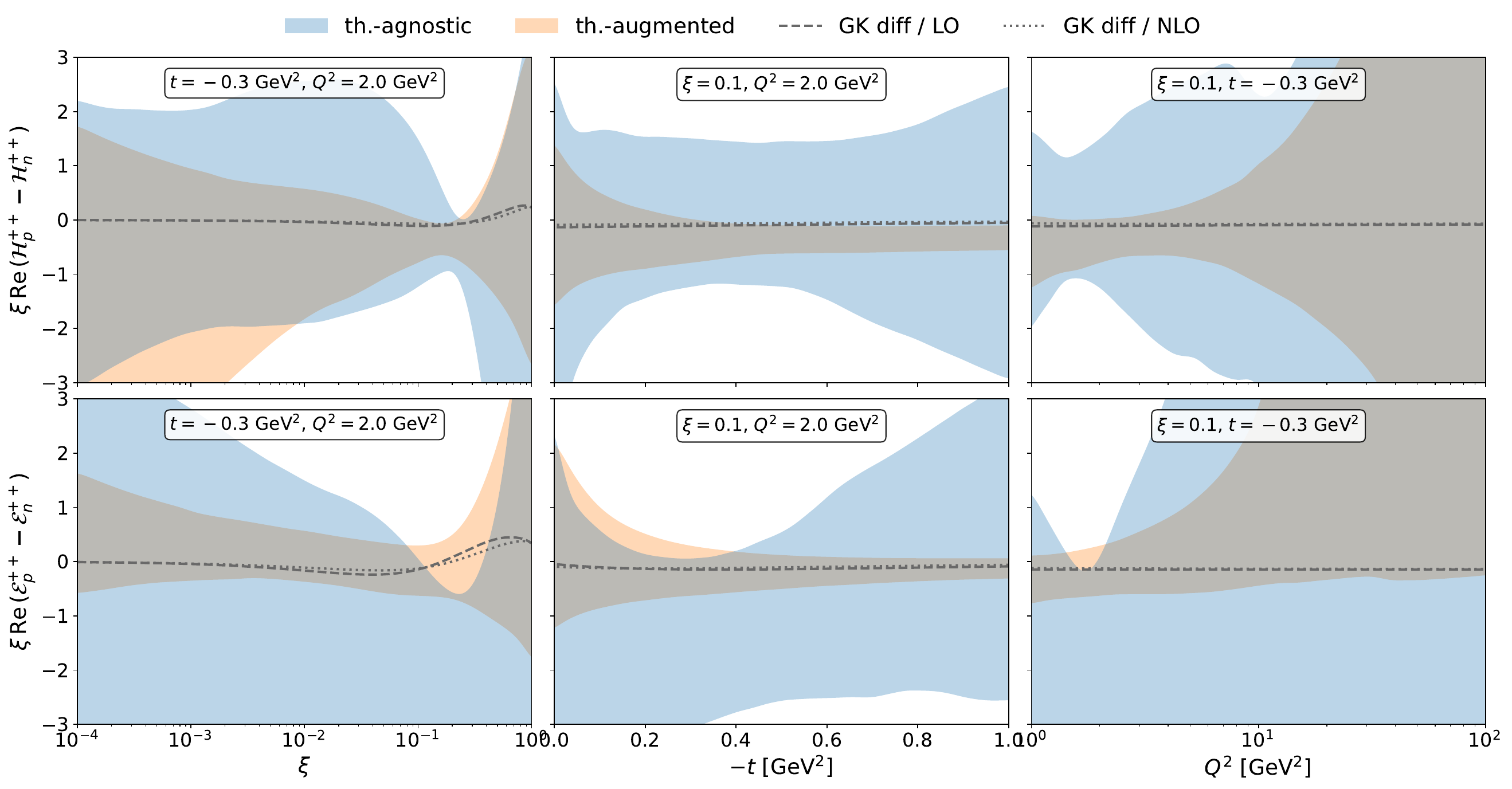}
    \includegraphics[width=1\textwidth]{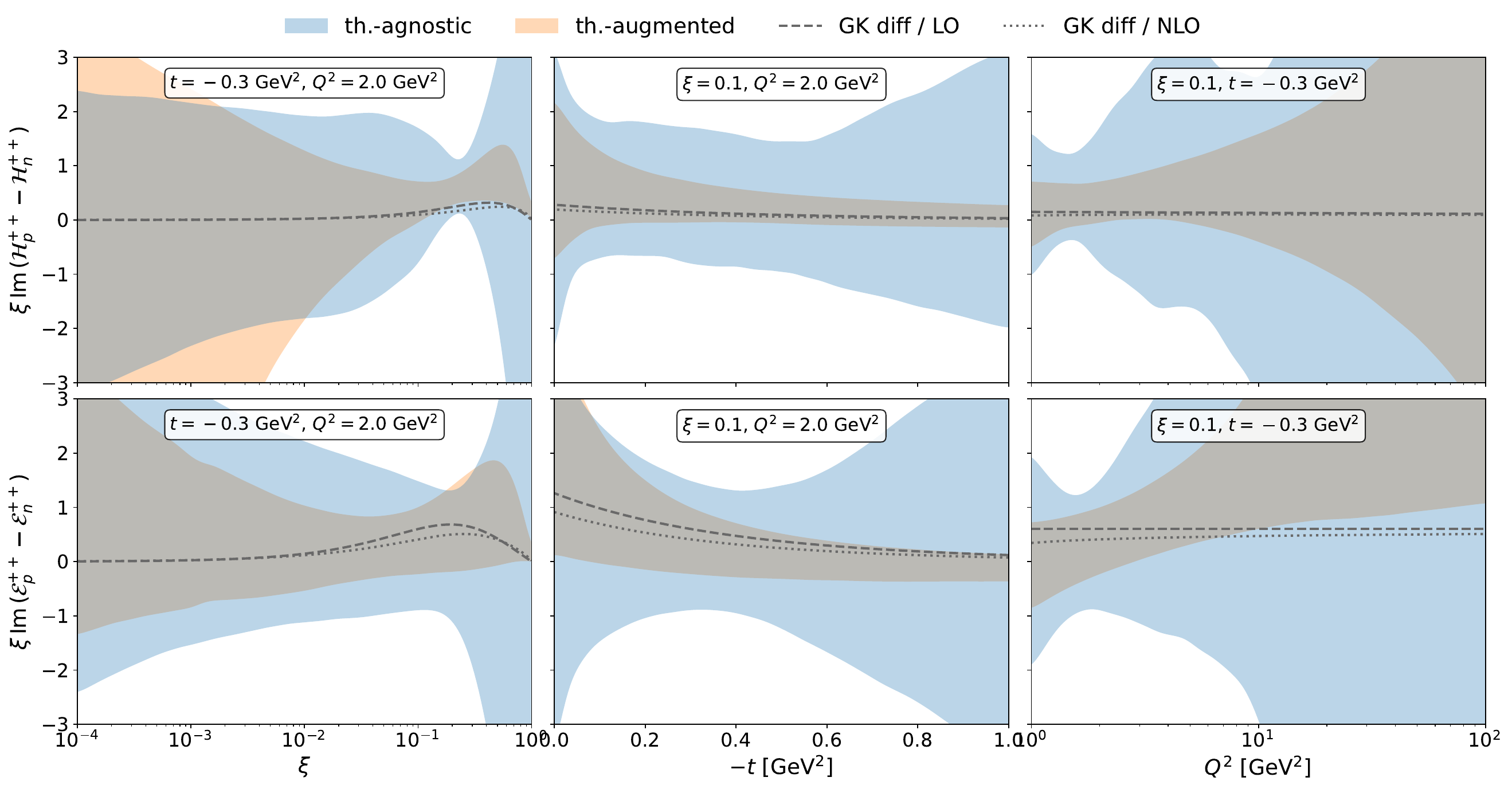}
    \caption{The difference between the proton and neutron CFFs (the isovector combination). For a further description, see the caption of Fig.~\ref{fig:result:cff_re_0}.}
    \label{fig:result:proton_neutron}
\end{figure}

The results for both models are presented in Fig.~\ref{fig:result:proton_neutron}. We observe that in the narrow kinematic region where the proton and neutron data overlap, a negative $\mathrm{Re}(\mathcal{H}^{++}_{p} - \mathcal{H}^{++}_{n})$ is favoured, although the effect is relatively weak. For the imaginary parts, there is a preference for positive $\mathrm{Im}(\mathcal{H}^{++}_{p} - \mathcal{H}^{++}_{n})$ and $\mathrm{Im}(\mathcal{E}^{++}_{p} - \mathcal{E}^{++}_{n})$, implying that $\mathcal{H}^{++}_{u} > \mathcal{H}^{++}_{d}$ and $\mathcal{E}^{++}_{u} > \mathcal{E}^{++}_{d}$. These observations are compatible with the KM analysis included in Ref.~\cite{CLAS:2024qhy}.
\section{Conclusions}
\label{sec:conclusions}

In this article, we have presented a new global analysis of DVCS data for both proton and neutron targets. Utilizing a newly developed framework within the PARTONS ecosystem, we have extracted CFFs, which are the most basic GPD-related observables one can unambiguously extract from experimental data. We deployed two distinct modeling strategies: a theory-agnostic approach and a theory-augmented model that rigorously enforces dispersion relations. Because extracting CFFs avoids the immediate pitfalls of the GPD deconvolution problem, our results provide a robust, computationally efficient distillation of the experimental data collected over the last two decades.

Our analysis highlights the pressing need to move beyond leading-order and leading-twist approximations in DVCS phenomenology. By explicitly incorporating the kinematically suppressed helicity-flip amplitudes, we observed a potentially non-zero, negative signal for $\mathcal{H}^{0+}$ in the high-$\xi$ regime. This behaviour points to the increasing relevance of higher-twist effects and signifies a breakdown of the strict leading-twist approximation. To put this result into more context, our global analysis favours the higher-twist scenario discussed in  Ref.~\cite{Defurne:2017paw} to explain JLab HallA data.

Furthermore, comparisons of our extracted dominant amplitudes with the GK model suggest that NLO corrections are crucial for an accurate description of the data. Although model-dependent, the significant impact of the NLO corrections observed in our analysis is consistent with earlier findings regarding the importance of higher-order effects in both DVCS phenomenology and combined DVCS/DVMP analyses~\cite{Moutarde:2013qs,Cuic:2023mki}.

The study is complemented by a direct comparison of the phenomenologically extracted DVCS subtraction constant with lattice-QCD predictions of the energy-momentum tensor (EMT) form factors. While the lattice results carry inherent systematic uncertainties, such as those stemming from unphysical pion masses and single-ensemble limitations, the theoretical predictions are compatible with our extraction, though the latter is characterised by large uncertainties, dominantly caused by a lack of experimental data sensitive to the real parts of CFFs. This feature confirms the results of previous studies~\cite{Kumericki:2019ddg,Dutrieux:2021nlz} based on JLab6 data.

We also conducted a test of GPD universality by projecting our extracted DVCS amplitudes to TCS. The resulting predictions exhibit good agreement with existing CLAS TCS measurements, underscoring the future potential of TCS data to constrain global GPD fits. Furthermore, the current TCS data already indicate that an NLO description is favoured.

Additionally, by extracting CFFs for both proton and neutron targets, we evaluated the isovector CFF combinations. This subtraction provides a clean, gluon-free probe of valence quark dynamics, indicating a preference for $\mathrm{Im}(\mathcal{H}^{++}_{u}) > \mathrm{Im}(\mathcal{H}^{++}_{d})$ and $\mathrm{Im}(\mathcal{E}^{++}_{u}) > \mathrm{Im}(\mathcal{E}^{++}_{d})$. These features are consistent with both the phenomenological extraction presented in Ref.~\cite{CLAS:2024qhy} and recent lattice-QCD computations presented in Ref.~\cite{Dutrieux:2026grg}.

Looking ahead, the CFFs extracted in this work serve as a foundational baseline for future GPD impact studies. Subsequent efforts will focus on tackling the explicit GPD deconvolution problem to perform a true 3D tomographic extraction. To achieve this, it will be essential to systematically incorporate next-to-next-to-leading order (NNLO) and higher-twist theoretical constraints. As lattice QCD and experimental precision continue to improve synchronously, the framework established here ensures that phenomenological analyses are prepared to propagate these advancements rigorously.

\backmatter

\bmhead{Acknowledgements}

We thank Silvia Niccolai for valuable discussions and help with the compilation of data used in this study. We also thank Victor Martinez-Fernandez for valuable discussions. This research was funded in part by the National Science Centre, Poland (grant SONATA BIS-15 No.~2025/58/E/ST2/00045 and grant IMPRESS-U No. 2024/06/Y/ST2/00155) and in part by the by l’Agence Nationale de la Recherche (ANR), project ANR-23-CE31-0019. For the purpose of Open Access, the author has applied a CC-BY public copyright licence to any Author Accepted Manuscript (AAM) version arising from this submission.
This work was made possible by Institut Pascal at Université Paris-Saclay with the support of the program “Investissements d’avenir” ANR-11-IDEX-0003-01.

\bibliography{Bibliography}

\end{document}